\documentclass[final,10pt,3p]{elsarticle}
\journal{Computer Physics Communications}
\usepackage{nicefrac}

\usepackage{amsmath,amssymb,amsfonts}
\usepackage{mathtools}
\usepackage{lmodern}
\usepackage{bm}
\usepackage{mathrsfs}
\usepackage{braket}
 
\usepackage{graphicx}
 
\usepackage{array}
\usepackage{dcolumn}
\usepackage{multirow}
\usepackage{tabularx}
 
\usepackage{algorithm}
\usepackage{algpseudocode}
 
\usepackage{csquotes}
\usepackage{enumitem}
\usepackage[normalem]{ulem}
\usepackage{todonotes}

\usepackage[colorlinks=true,
citecolor=blue,
linkcolor=red,
urlcolor=teal]{hyperref}
\usepackage{upgreek}
\newcommand{\muv}{\boldsymbol{\upmu}}
\newcommand{\thetav}{\boldsymbol{\uptheta}}
\newcommand{\etav}{\boldsymbol{\upeta}}
\newcommand{\sigmav}{\boldsymbol{\upsigma}}
\newcommand{\omegav}{\boldsymbol{\upomega}}
\newcommand{\tauv}{\boldsymbol{\uptau}}

\newcommand{\xv}{\mathbf{x}}
\newcommand{\cv}{\mathbf{c}}
\newcommand{\bv}{\mathbf{b}}

\newcommand{\gv}{\mathbf{g}}
\newcommand{\zv}{\mathbf{z}}
 
\begin{document}
 
\begin{frontmatter}
 
\title{Dictionary learning approach to analytic continuation of imaginary-time quantum Monte Carlo data}
 
\author[casus,hzdr]{Thomas Chuna\corref{cor1}}
\cortext[cor1]{Corresponding author}
\ead{t.chuna@hzdr.de}

\author[casus,tud]{Phil-Alexander Hofmann}
 
\author[hzdr,tud]{Alexander Benedix Robles}
 
\author[hzdr,casus]{Tobias Dornheim}
 
\address[casus]{Center for Advanced Systems Understanding (CASUS), Helmholtz-Zentrum Dresden-Rossendorf (HZDR), D-02826 G\"orlitz, Germany}

\address[hzdr]{Institute of Radiation Physics, Helmholtz-Zentrum Dresden-Rossendorf (HZDR), D-01328 Dresden, Germany}

\address[tud]{Technische Universit\"at Dresden, D-01062 Dresden, Germany}

\begin{abstract}
Many fields of physics use quantum Monte Carlo (QMC) simulations to simulate quantum systems in imaginary-time $\tau$ and estimate imaginary-time correlation functions (ITCF). However, extracting dynamic $\omega$-dependent quantities from ITCFs is a notoriously difficult task, known as analytic continuation (AC), that amounts to solving an exponentially ill-conditioned inverse problem. Within the AC literature, there are competing stochastic and regularized approaches, as well as an emerging collection of works using parameterized models like neural networks. Here we transcend the traditional divides between the communities, introducing the regularized stochastic optimization method (RSOM). This method reformulates AC as a dictionary learning problem, discovering a sparse dictionary to represent the solution. Our approach is motivated by the astounding results dictionary learning has produced in many scientific fields. Remarkably, RSOM discovers a sparse dictionary that maps a high-dimensional ill-conditioned inverse problem to a low-dimensional problem. We demonstrate that the method yields competitive results for common synthetic test problems as well as for authentic QMC data from the finite temperature electron gas. Due to the low dimensionality, there is reasonably finite uncertainty in the dictionary, which is leveraged to create an \textit{ad hoc} post-processing estimate of the dictionary coefficients' sensitivity to perturbations. This work exposes that a dictionary exists within all stochastic and regularized methods and that dictionary learning provides a new angle of attack for future AC methods.
\end{abstract}
 
\end{frontmatter}



\section{Introduction}
The analytic continuation (AC) problem is a long-standing grand challenge to the quantum Monte Carlo (QMC) community. QMC methods are used in quantum chemistry, materials science~\cite{Foulkes_RMP_2001, anderson_QMCtextbook_2007, Ceperley_RevModPhys_1995}, high energy physics~\cite{gattringer_LQCDtextbook_2009, knechtli_LQCDtextbook_2017}, nuclear structure~\cite{Leidemann_PPNP_2013}, and Higgs models~\cite{DelDebbio_EuroPhysC_2023}, but the applications of QMC simulations are limited because simulations generate estimates of $N$-body imaginary-time correlation functions (ITCF), denoted $F(\tau)$, rather than real-time correlations. Thus, the ITCFs are related to dynamic properties, denoted by spectral function $S(\omega)$, by an analytic continuation rather than the Fourier transform~\cite{Jarrell_PhysRep_1996}. Essentially, the AC problem is determining dynamic spectrum $S(q, \omega)$, given ITCF data $F(q, \tau)$ and transformation $\mathcal{A}(\tau,\omega)$, which are related by~\cite{buividovich_arXiv_2025, Tripolt_CPC_2019, Dornheim_JCP_ITCF_2021, Lovato_PRC_2015,  Hansen_PRD-GBG_2019, Filinov_PRB_2023, Parnes_PRL_2026, rothkopf_FrontiersPhys_2022}
\begin{align}\label{eq:ACproblem_continuous}
    F(q, \tau) = \int_{- \infty}^{\infty}  d\omega \,  \mathcal{A}(\tau,\omega) S(q, \omega) \, .
\end{align}

In practice, inverse problem \eqref{eq:ACproblem_continuous} is solved numerically. The finite periodic space and imaginary-time boundaries used in QMC simulations mean that $F(q, \tau)$ is known on a discrete grid of $N_q \times N_\tau$ points. For a given $q_0$ value, the continuous $\omega$ variable is discretized to $N_\omega$ points in order to formulate \eqref{eq:ACproblem_continuous} as a matrix inversion. This inversion is conducted for each $q_0$ separately. We distinguish the discrete inverse problem formulated by discretizing the $\omega$-domain of the continuous AC problem \eqref{eq:ACproblem_continuous},
\begin{align}\label{eq:ACproblem_discrete}
    \bv = A \xv \, ,
\end{align}
with $A_{ij} = \Delta \omega  \, \mathcal{A}(\tau_i,\omega_j)$, $x_j = S(q_0, \omega_j)$, and $b_i = F(q_0, \tau_i)$.  Typically $N_\omega > N_\tau$ and the condition number of the matrix $\kappa(A)$ is infinity. Thus, \eqref{eq:ACproblem_discrete} is under-determined and ill-conditioned, so direct inversion is futile; for more discussion on these points see \cite{shi_CPC_2023}.

Methodological divides have emerged about how to solve this inverse problem. Firstly, regularized optimization solves for $\xv$ via a single walker's gradient-informed optimization of a chi-square goodness-of-fit metric $\chi^2[\xv\mid \bv]=\frac{1}{2} \Vert A \xv - \bv \Vert^2_{C^{-1}}$ combined with a regularization term $r[ \cdot ]$ weighted by $\lambda$, which may or may not rely on a Bayesian prior $\muv$~\cite{Jarrell_PhysRep_1996, Asakawa_PPNP_2001, BurnierRothkopf_PRL_2013, Otsuki_PRE_2017,  Otsuki_JPSJ_2020, prokofev_JETP_2013, Han_PRB_2022, chuna_JPA_2025, robles_CPC_2025}. This is expressed as
\begin{align} \label{eq:GLSEntropy}
    S = \max_\xv \, \,  -\chi^2[\xv\mid \bv] 
    + \lambda r[ \xv \mid \muv] \, .
\end{align}
Alternatively, stochastic approaches typically remove the regularization term, (\textit{i.e.}, $\lambda \rightarrow 0$), release one or more walkers that obtain representative solutions via Metropolis-Hastings sampling, and then estimate $x$ by averaging the solutions~\cite{Sandvik_PRB_1998, Mishchenko_PRB_2000, Vitali_PRB_2010, Saccani_Supersolid_PRL_2012, prokofev_JETP_2013,shu_arXiv_2015, bao_PRB_2016, Nichols_PRE_2022, ShaoSandvik_PhysRep_2023}. However, as argued by Goulko \textit{et al}.~\cite{Goulko_PRB_2017}, as well as Bergeron and Tremblay~\cite{bergeron_PRE_2016}, neither optimization is conclusively better.

An alternative third approach exists, described by Shao and Sandvik~\cite{ShaoSandvik_PhysRep_2023} as, ``a very good solution to the analytic continuation problem would be to just use $\chi^2$ minimization with a suitable functional form... However, it is very difficult to construct a generic form with sufficient flexibility...''. This parameterized model approach comes in one of two kinds: functional forms constrained to only a few free parameters~\cite{Sandvik_PRL_2001, Katz_PRB_2014, buividovich_arXiv_2024, groth_PRB_2019, Filinov_PRB_2023, Efremkin_PRL_2026} or neural networks~\cite{Yoon_PRB_2018, Kades_PRD_2020, Fournier_PRL_2020,  xie_DCDSS-ABNueralNetwork_2021, Chen_PoS_2021, Huang_PRB_2022, wang_PRD_2022, wang_JPhys_2023}. While explicit functional forms reduce the optimization space, they have insufficient flexibility. On the other hand, neural networks are universal function approximators, but they introduce many parameters, expanding the optimization space and requiring training data. A balance can be struck between these approaches; auto-encoders provide a data-driven low-dimensional parameterization of the solution. In recent work by Meir \textit{et al}.~\cite{Meir_arXiv_2026}, it has been demonstrated the neural network solution space is parsimonious, \textit{i.e.}, admits a low dimensional representation. This motivates the use of encoders and other flexible low-dimensional parameterizations, but they are scarcely used in the AC literature~\cite{Chen_PoS_2021, Rumetshofer_PRB_2019, Zhao_PRB_2026}.

In this work, we transcend the divide between these three approaches, creating the regularized stochastic optimization method (RSOM). In the problems we consider, the RSOM discovers a low-dimensional Gaussian parameterization of the solution (\textit{i.e.},  understood as a sparse dictionary or encoder) via dictionary learning. This formulates the continuous ill-posed AC problem as an optimization over the dictionary and the dictionary code, given later in \eqref{eq:DLformulation}. The RSOM's dictionary learning process scans over the number of Gaussians and stochastically optimizes the Gaussian widths and centers in order to ensure the dictionary achieves an acceptable goodness-of-fit. In practice, the RSOM can reconstruct multi-peak structures without sacrificing performance on smoother structures. Due to the low dimensionality, the dictionary (for a given number of kernels) has a reasonably finite condition number and uncertainty, which we leverage to approximate the maximum error in the Gaussian coefficients', though this bound does not account for the uncertainty in the number of kernels or choice of kernels.

The primary impact of this work is presenting the AC as a dictionary learning problem, rather than as an ill-conditioned inversion problem. We do this because the AC literature has struggled for three decades to move beyond the seminal maximum entropy method~\cite{Jarrell_PhysRep_1996}, while sparse dictionary learning has produced many useful results in dynamical systems and image processing (\textit{e.g.}, computer vision and medical imaging). This formulation indicates that a dictionary exists within all stochastic and regularized methods and that typically this dictionary is fixed \textit{a priori} without respect to the data.

This paper is organized as follows: In section~\ref{sec:theory} we formulate the RSOM. In section~\ref{sec:results}, we present results for the RSOM on synthetic inverse problems comparing to other leading methods. This compares its performance and demonstrates generality across problem types. In section~\ref{sec:UEG}, we apply the method to authentic QMC data obtained for the uniform electron gas. In section~\ref{sec:conditioning}, we investigate the sensitivity of the Gaussian coefficients' to noise and propose an error bound on the coefficients'. Finally, in section~\ref{sec:conclusions} we summarize our results, draw conclusions, and plot a course for future work. We provide the details of the RSOM's implementation in~\ref{app:implementation}.

\section{Numeric Formulation}\label{sec:theory}
The RSOM separates the AC problem into two nested pieces: The inner piece, which we refer to as the \textit{discovered discrete inverse problem}, takes a set of Gaussians and fits their coefficients to the data using non-negative least squares optimization (NNLS). The outer piece selects the number of Gaussians as well as their centers and widths via a dimensionality scan and stochastic sampling to ensure the inner piece achieves a reasonable goodness-of-fit. In this section, we formulate: the dictionary in subsection \ref{sec:kernelformulation}, the inner piece in subsection \ref{sec:c_formulation}, and the outer piece in subsection \ref{sec:DLformulation}. Finally, we discuss how the RSOM is related to prior publications in subsection \ref{sec:priorwork}.

\subsection{Representing the solution as a dictionary}\label{sec:kernelformulation}
To start, we assume a fixed $q_0$ and neglect the $q_0$ dependence in $S(q_0,\omega)$ and represent it as a linear combination of $N_c$ continuous kernel functions
\begin{align}\label{eq:kernel_parameterization}
    S(\omega) = \sum_j^{N_c} c_j \mathcal{K}_j(\omega) \, ,
\end{align}
where the set of kernel functions  $\{\mathcal{K}_0(\omega), \ldots \}$ is referred to as a dictionary. Here the $j$th kernel may have one or many parameters. For example, a Dirac delta has one parameter for its center $\omega_j$, leading to $\mathcal{K}_j(\omega) = \delta(\omega - \omega_j)$, while an unnormalized Gaussian kernel has two parameters for its center $z_j$ and width  $\sigma_j$ so that $\mathcal{K}_j(\omega) = \exp[-(\omega - z_j)^2/\sigma^2_j]$. We denote the collection of all parameters across all $j$ as $\thetav$. Substituting \eqref{eq:kernel_parameterization} into \eqref{eq:ACproblem_continuous} produces
\begin{subequations}  \label{eq:reformulatedAC}
\begin{align}
    F(\tau) &= \sum_j^{N_c} c_j \mathcal{R}_j(\tau) \, ,
    \\ \mathcal{R}_j(\tau) &= \int_{-\infty}^\infty d\omega \, \mathcal{A}(\tau,\omega) \mathcal{K}_j(\omega) \, . \label{eq:kerneltransform}
\end{align}
\end{subequations}
The application of $\mathcal{A}$ to $\mathcal{K}_j$ converts the kernel from $\omega$ to $\tau$ space. 

To numerically solve \eqref{eq:reformulatedAC} for $\cv$, we discretize $\tau$ to obtain matrix $R_{ij} = \mathcal{R}_j(\tau_i)$. This leads to the discrete formulation,
\begin{align}\label{eq:reformulatedACproblem_discrete}
    \bv = R \, \cv \, ,
\end{align}
This form has the nice property that the solution $S(\omega)$ obtained from inserting $\cv$ into \eqref{eq:kernel_parameterization} is $\omega$-grid independent. However, if an analytic form of the transformed kernels $\mathcal{R}_j(\tau)$ is not known, as is generally the case, then the regression matrix can be computed via numeric integration as $R_{ij} = \sum_k A_{ik} \mathcal{K}_j(\omega_k)$, which is conveniently expressed in matrix form as,
\begin{align}
    R = A \, K.
\end{align}
Here matrix $A$ is the $N_\tau \times N_\omega$ discretization of transformation $\mathcal{A}$ and matrix $K$ is the $N_\omega \times N_c$ discretization of the dictionary, so the $j$th column of $K$ contains the values of $j$th kernel function evaluated on the discrete $\omega$-grid. 

We note that often the system's detailed balance condition is exploited to modify the integration domain as $R_{ij} = \int_0^\infty  \mathcal{A}(\tau_i,\omega) \left( \mathcal{K}_j(\omega) + e^{\beta \omega} \mathcal{K}_j(-\omega) \right) \, d\omega $~\cite{chuna_JPA_2025, robles_CPC_2025}, where temperature is $T = 1/\beta$. This is beneficial because it reduces to the positive $\omega$ domain ergo reducing the size of the inverse problem. If this is the case, then the matrix $A$ will be modified. For the Laplace transform, this leads to $A_{ij} = e^{-\tau_i \omega_j} + e^{-(\beta - \tau_i) \omega_j}$. Thus, at $T=0$, we do a Laplace transform over the positive $\omega$ domain. We assume the reader is familiar with how this is done for their specific AC problems.

\subsection{Fitting the dictionary code to the data}\label{sec:c_formulation}
In this subsection we present the inner discovered discrete inverse problem, \textit{i.e.}, stating the problem for inverting $R$ to solve for $\cv$ in \eqref{eq:reformulatedACproblem_discrete}. In contrast to the last subsection, we make the dependence of the dictionary $K$ on the number of kernels $N_c$ and the parameters of those kernels $\thetav$ explicit by presenting the dictionary as $K(N_c,\thetav)$. For a given dictionary $K(N_c,\thetav)$, chi-square goodness-of-fit for the kernel coefficients $\cv$ is given
\begin{subequations}
\label{eq:PyLITcostfunction}
\begin{align}
    S[ K(N_c, \thetav) ] &= \min_{\cv \ge 0} \chi^2[K(N_c, \thetav ), \, \cv] \, ,
\end{align}
where $c \in \mathbb{R}^{+,N_c}$ is often referred to as the dictionary code, and 
\begin{align} \label{eq:chi-sq_Npc}
    \chi^2[K(N_c,\thetav), \, \cv] &= \Vert F - A K(N_c, \thetav ) \, \cv \Vert^2_{C^{-1}} 
\end{align}
\end{subequations}
Equation \eqref{eq:PyLITcostfunction} can be obtained from \eqref{eq:GLSEntropy} by replacing $x$ with the discrete version of \eqref{eq:kernel_parameterization}, \textit{i.e.}, $\xv = K(N_c,\thetav) \, \cv$. Constraining $\cv \geq 0$, along with the constraint the kernels are non-negative, ensures that the solution is non-negative. 

Dictionary $K(N_c,\thetav)$, while explicit here, is also present in the typical formulation \eqref{eq:GLSEntropy}. There we have a dictionary $K(N_c=N_\omega,\thetav)$ of $N_\omega$ Dirac deltas, with $\thetav$ defining the $\omega$-grid locations and the coefficients $\cv$ defining the solution $x$. Thus, \eqref{eq:PyLITcostfunction} has exposed a choice is always being made and we are condemned to freedom. Additionally, in the case where a sparse $K(N_c,\thetav)$ is identified, then we can interpret this $K(N_c,\thetav)$ as an encoder because the dimension of the optimization problem is reduced from $\dim(\xv)= N_\omega$ to $\dim(\cv)=N_c$. So the typical formulation is not leveraging the data to construct an encoder and leaving it as a user input, though the discretization is typically leveraged to reduce the cost of the inverse problem~\cite{bergeron_PRE_2016}, from the perspective of the RSOM, the dictionary can also be leveraged to regularize and improve the estimate. This aligns with Ghanem \textit{et al}.~\cite{Ghanem_PRB_2020}. As we shall see, if many Gaussian functions are needed to represent a function the optimization of kernel coefficients becomes ill-conditioned.

\subsection{Constructing the dictionary from the data} \label{sec:DLformulation}
Now that we have exposed that a dictionary is always present in the inverse problem, we must construct the dictionary from the data using dictionary learning (DL). This is formulated as
\begin{align}\label{eq:DLformulation}
    S = \underset{N_c , \,  \thetav}{\text{DL}} \left( \min_{\cv \ge 0} \chi^2[K(N_c,\thetav), \, \cv] \right).
\end{align}
Constructing the dictionary from the data rather than fixing it \textit{a priori} ensures RSOM controls the bias of its solution. 

In practice, we split the dictionary learning of $K(N_c,\thetav)$ into two steps. First, a dimensionality scan over $N_c$ to select the sparsest dictionary needed to represent the data~\cite[Chapter 4]{brunton_textbook_2022}. Restricting the solution to a few Gaussians removes most admissible spectra, but such parsimony is arguably the ultimate regularization of an optimization problem~\cite{kutz_NonlinearDynamics_2022}. Second, a stochastic optimization of the Gaussian parameters $\thetav$, where we use stochastic approaches because it makes the solutions more robust to noise with increasing $N_c$ and allows post-processing assessments of the uncertainty in the matrix $R$. Together these steps form a parsimony \textit{regularized stochastic optimization method}. This leads to:
\begin{align}\label{eq:RSOMopt}
    S = \underset{N_c}{\text{scan}} \left[ \min_{\thetav} \left( \min_{\cv \ge 0} \chi^2[K(N_c,\thetav), \, \cv] \right) \right].
\end{align}
Equation \eqref{eq:RSOMopt}, is our particular implementation of \eqref{eq:DLformulation}, but many different algorithms can be linked together to select $N_c$, $\thetav$, and $\cv$. 

Since the dimensionality scan approach seeks sparsity, then \eqref{eq:RSOMopt} can be understood as regularizing \eqref{eq:PyLITcostfunction} by the notorious $L_0$ regularizer, \textit{i.e.}, $\Vert \cv \Vert_0$. This problem is NP hard~\cite{natarajan_Computing_1995} and has no best approach. So the reformulation has traded one difficult problem for another. However, the field of sparse dictionary learning has developed many impactful tools.

Drawing from the field of dictionary learning, the implementation of \eqref{eq:RSOMopt} and the reasoning behind that implementation are discussed in~\ref{app:implementation}. Summarizing the Appendix, the minimization over $\cv$ is handled by a standard NNLS optimization, the minimization of $\thetav$ is handled by a gradient-informed Metropolis-Hastings algorithm, and the dimensionality scan seeks to identify the sparsest dictionary by scanning over the number of kernels to identify the smallest $N_c$ where the goodness-of-fit is good enough. The uncertainty in $S$ is estimated by computing the standard error from the distribution of acceptable solutions at a given $N_c$.

In summary, \eqref{eq:DLformulation} is the primary result of the paper because it formulates the AC problem as a dictionary learning problem. This formulation is motivated by the extraordinary results that dictionary learning has produced in other fields and supported by independent observations of a parsimonious AC solution~\cite{Meir_arXiv_2026, Zhao_PRB_2026}.

\subsection{Relation of the RSOM to prior analytic continuation work}\label{sec:priorwork}
Equation \eqref{eq:kernel_parameterization} constitutes a dictionary that maps the typical optimization problem from a $N_\omega$-dimensional space to a smaller $N_c$-dimensional space. This is like Bryan's well-known maximum entropy method (MEM)~\cite{bryan_EuroBiophys_1990, Asakawa_PPNP_2001}, which represents the solution in terms of $A$'s singular vectors $U_j(\omega_k)$, linearly parameterizing the solution in terms of variable $\cv \in N_c$ in the low-dimensional singular subspace as
\begin{align}\label{eq:Bryan_parameterization}
    S(\omega_k) = \mu(\omega_k) \exp\left(\sum_j^{N_c} c_j \, U_j(\omega_k) \right) \, .
\end{align}
However, this SVD-based mapping is fixed \textit{a priori} by the matrix $A$ in \eqref{eq:ACproblem_discrete}, without respect to the data $\bv$ or the solution $\xv$, so it reduces to a subspace that can be too small~\cite{rothkopf_Data_2020}. In practice, Bryan's parameterization works well for smooth, single peak spectra, but not sharp multi-peaked spectra~\cite{Fischer_PRD-SmoothedBRM_2018}. In contrast, Gaussian mixtures are universal function approximators~\cite{McLachlan_textbook_1988}. This means for small $N_c$ there is no guarantee that the data can be well-represented, but with a sufficient number of Gaussians any reasonable function can be represented, \textit{i.e.}, achieve a reasonably small goodness-of-fit.

RSOM's use of a non-linear parameterization \eqref{eq:kernel_parameterization} is similar in spirit to Rothkopf and collaborators' ``extended MEM''~\cite{rothkopf_JCP_2012, kelly_PRD_2018}, which replaces the singular vector basis with non-linear (sinusoidal) functions.
\begin{subequations}
\begin{align}\label{eq:Rothkopf_parameterization}
    S(\omega_k) &= \mu(\omega_k) \exp\left(c_1 + \sum_{j=1}^{N_c} \left\{ c_{2j} \sin(\omega_k) + c_{2j+1} \cos(\omega_k) \right\}\right) \, ,
\end{align}
\end{subequations}
However, the extended MEM is known to produce spurious peaks at large $\omega$. The RSOM does not exhibit this behavior because Gaussian kernels are local in $\omega$-space and RSOM searches for the fewest possible peaks needed to describe the data, whereas extended MEM used a fixed 50 sinusoidals.

RSOM’s dictionary expresses the solution as a linear combination of a few functions.
In modern AC literature, Rumetshofer \textit{et al}.~\cite{Rumetshofer_PRB_2019} parameterizes the solution in terms of a few Lorentzians and then fits the Lorentzians' coefficients and parameters with an entropic regularized fidelity. Lorentzian kernels are well motivated for physical systems with sharp quasi-particle poles, but their heavy tails lead to undefined moments and an ill-defined Laplace transform~\cite{blitzstein2019introduction}. For the general AC problem, where sum rules assess the quality of a solution~\cite{chuna_PRB_2025}, Gaussian kernels, which have well-defined moments, are more appropriate. 

Gaussian kernels have been explored by Robles \textit{et al}.'s Python Laplace Inverse Transform (PyLIT) code~\cite{robles_CPC_2025}. However, the kernel parameters are selected \textit{a priori} by fitting to a Bayesian prior, rather than to the data. This introduces a bias~\cite{robles_CPC_2025, chuna_arXiv_2026}, which PyLIT mitigates by including many kernels in the dictionary to expand the space of representable solutions. With each additional kernel, the problem \eqref{eq:PyLITcostfunction} becomes more ill-conditioned and thus regularization is needed. The RSOM addresses both of the issues that plague PyLIT by using the data, instead of a prior, to create a low-dimension Gaussian dictionary.

The RSOM shares a desire for sparsity that is common in the AC literature. Otsuki \textit{et al}.~\cite{Otsuki_PRE_2017, Otsuki_JPSJ_2020} regularize the typical AC formulation with the $L_1$-norm, so their approach is essentially pruning a dense dictionary of Dirac deltas. However, as noted by Huang and Yang~\cite[section II.C]{Huang_PRB_2022}, representing the QMC data using a combination of exponentials (the Laplace transform of Dirac deltas) introduces a fundamental limitation on the quality of the solution. Huang and Yang overcome this barrier by replacing the dense dictionary with a neural network (NN) trained with supervised-learning and again regularized by the $L_1$ norm. In contrast to these approaches, the RSOM enforces sparsity through the $L_0$ constraint and builds a sparse dictionary, rather than pruning a dense dictionary. As a result, the RSOM dictionary has a smaller condition number and a smaller dimensionality $N_c$, so no training set is needed. This approach resembles Maltz's parsimonious exponential spectral analysis~\cite{maltz_PhysMedBio_2002}, which increases the number of exponential decays until the data is well represented.

In summary, the RSOM manifests the low-dimensional compression idea from Bryan, the non-linear Gaussian parameterization from PyLIT, the sparsity motivation from Otsuki \textit{et al}., and contributes a data-driven approach to discovering the appropriate parameterization/dictionary. Together, these ideas allow the RSOM to regularize via parsimony rather than via a Bayesian prior.

\section{Demonstrating performance of the RSOM on synthetic problems}\label{sec:results}

With the RSOM formulation established and discussed, we investigate its performance on problems with known solutions. The RSOM code base and these examples are freely available online~\cite{github_MEMcode}. Each subsection below presents a problem and the results.

\subsection{Synthetic data from Laplace transformed rho-meson spectrum}
\begin{figure}
    \centering
    \includegraphics[width=0.5\linewidth]{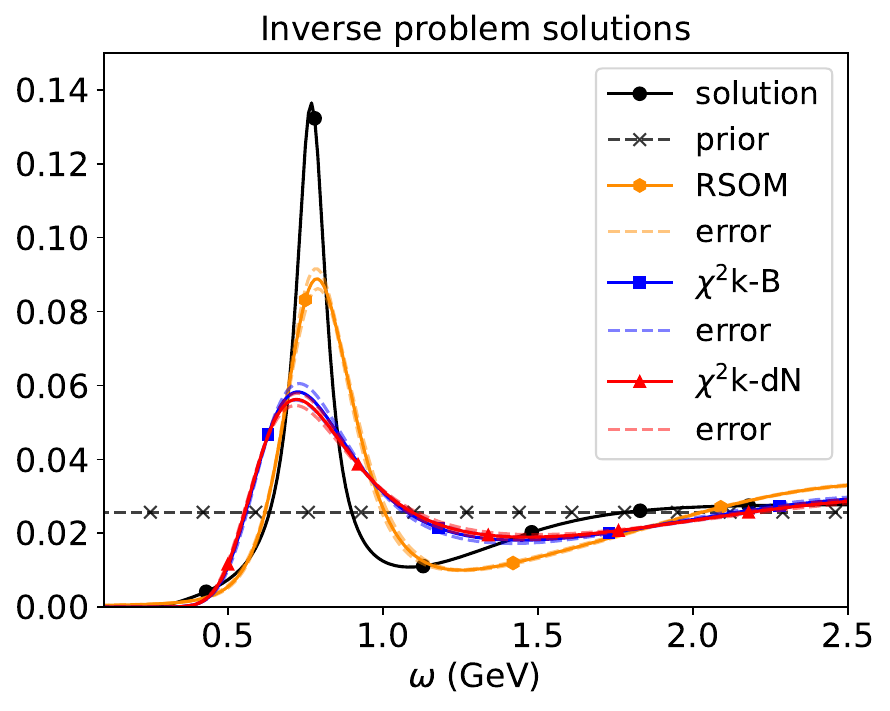}%
    \includegraphics[width=0.5\linewidth]{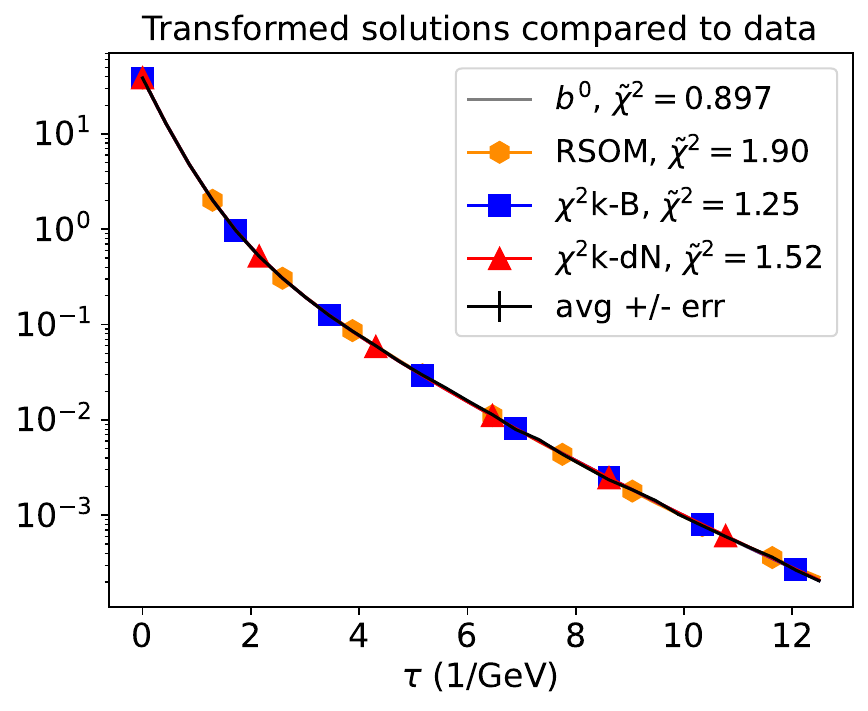}
    \caption{This plot demonstrates the RSOM's performance for the AC of a realistic rho-meson spectrum, comparing with the typical entropic methods which incorporate a Bayesian prior. Left: Solutions in $\omega$-space. Right: Transformed solutions in $\tau$-space with their associated goodness-of-fit to the data (\textit{i.e.}, the reduced chi-square).}
    \label{fig:rho-meson}
\end{figure}

We consider a test problem from lattice QCD~\cite{Asakawa_PPNP_2001} to connect to a broader audience. This problem has been investigated extensively~\cite{chuna_JPA_2025} and similar shapes are often studied~\cite{Ding_PRD_2012, Chen_PoS_2021, wang_PRD_2022}. The integral transform is defined by
\begin{align}
    \mathcal{A}(\tau,\omega)= \omega^2 e^{-\tau \omega} \, 
\end{align}
and the problem is formulated on uniform grids $\omega_j = j \, \Delta \omega$ with $j = 8, \dots, N_\omega$, $N_\omega=350$, and $\Delta \omega = 0.01$ GeV as well as $\tau_i = i \, \Delta\tau$ with $i=0,\dots,N_\tau-1$, $N_\tau=30$, and $\Delta\tau = 0.085 \,\mathrm{fm} \approx 0.431 \, \mathrm{GeV}^{-1}$. We remove the first 8 $\omega$ points because the spectral function is numerically zero here.  Using $A_{ij}= \Delta \omega \, T(\tau_i,\omega_j)$, we create a noiseless signal by transforming the true solution $\xv^0$, a $\rho$-meson spectral function,
\begin{align}
    x^0_k = \frac{2}{\pi}\left(\text{pole}(\omega_k)+\text{cut}(\omega_k)\right) \, .
\end{align}
Here the pole term arises from a $\rho$-meson bound state and is a Lorentzian centered at $m_\rho=0.77,\mathrm{GeV}$ with energy-dependent width. The complex cut term arises from the unbound states or ``continuum'' and is a logistic growth with an inflection point at $\omega_0=1.3$ GeV. Plots of $\xv^0$ and $\bv^0$ are given in Figure~\ref{fig:GaussianKernel} bottom left and right. Further details on the $\rho$-meson functional form are contained in~\cite{Shuryak_RevModPhys_1993, Asakawa_PPNP_2001, chuna_JPA_2025}. The data samples are generated by adding Gaussian noise to each component of the noiseless signal $\bv^0 = A \xv^0$,
\begin{align}
    b^s_i \sim \, b^0_i + \mathcal{N}\left(z = 0, \sigma = \sigma_0 \, b^0_i\left(\frac{\tau_i+\delta\tau}{\Delta\tau}\right) \right) \, ,
\end{align} 
where $\sigma_0 = 10^{-2}$ and $\delta \tau=0.1$. From $N_{\mathrm{s}}=100$ realizations, the sample mean defines $\bv$ and the empirical variance determines the noise level; this level of noise is too small to be distinguished within Figure~\ref{fig:rho-meson}. We create the regression matrix for the RSOM algorithm using matrix multiplication as $R_{ij} = \sum_k A_{ik} K_{jk}$, where $K_{jk}(\omega) = \exp[-(\omega_k - z_j)^2/\sigma^2_j]$. 

To assess the method, we compare RSOM's estimate to the common entropic least squares problem, often referred to as the maximum entropy method~\cite{Kaufmann_CPC-anacont_2023}. We solve the entropic least squares via Bryan's (B) approximate Levenberg-Marquardt optimizer~\cite{bryan_EuroBiophys_1990, Jarrell_PhysRep_1996, Asakawa_PPNP_2001, rothkopf_Data_2020} as well as the exact dual Newton (dN) optimizer~\cite{chuna_JPA_2025, barnfield_arXiv_2026}. For the entropic methods, instead of the typical Bayesian posterior averaging over the regularization weight~\cite{gull_MaxEntBayesianMethods_1989}, we use Kaufmann and Held's $\chi^2$-kink ($\chi^2$k) procedure~\cite{Kaufmann_CPC-anacont_2023} for the reasons argued therein. The RSOM's result as well as the entropic approaches are presented in Figure.~\ref{fig:rho-meson}.

We find that the RSOM's peak is closer to the true solution than the entropic methods. Other methods have found similar improvements with respect to peak reconstruction over Bryan's algorithm, but only when the noise is large and a limited number of basis vectors are used~\cite{Ding_PRD_2012, Chen_PoS_2021, wang_PRD_2022, Yoon_PRB_2018, xie_DCDSS-ABNueralNetwork_2021, Fournier_PRL_2020}. As shown by Chuna \textit{et al}.~\cite{chuna_PRR_2026}, Bryan's approximate algorithm breaks down in such a limit, thus such accomplishments are less convincing than what we demonstrate. Here we find that both the dual Newton and Bryan approach yield the same result, so we can conclude that the noise is sufficiently small and the number of basis vectors is sufficient for Bryan's algorithm to be valid.
 
\subsection{Synthetic data from Laplace transformed double Gaussian spectrum}
\begin{figure}
    \centering
    \includegraphics[width=0.5\linewidth]{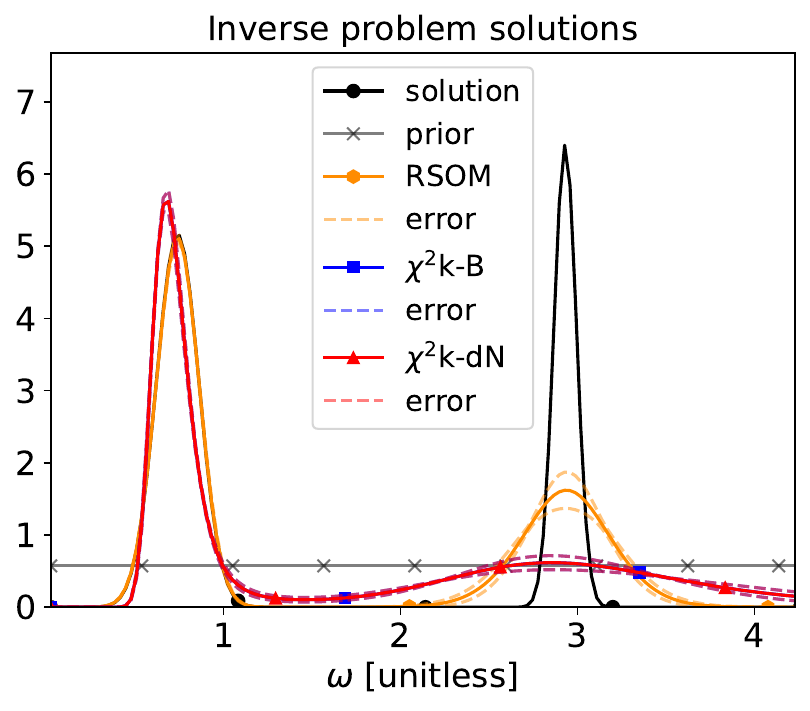}%
    \includegraphics[width=0.5\linewidth]{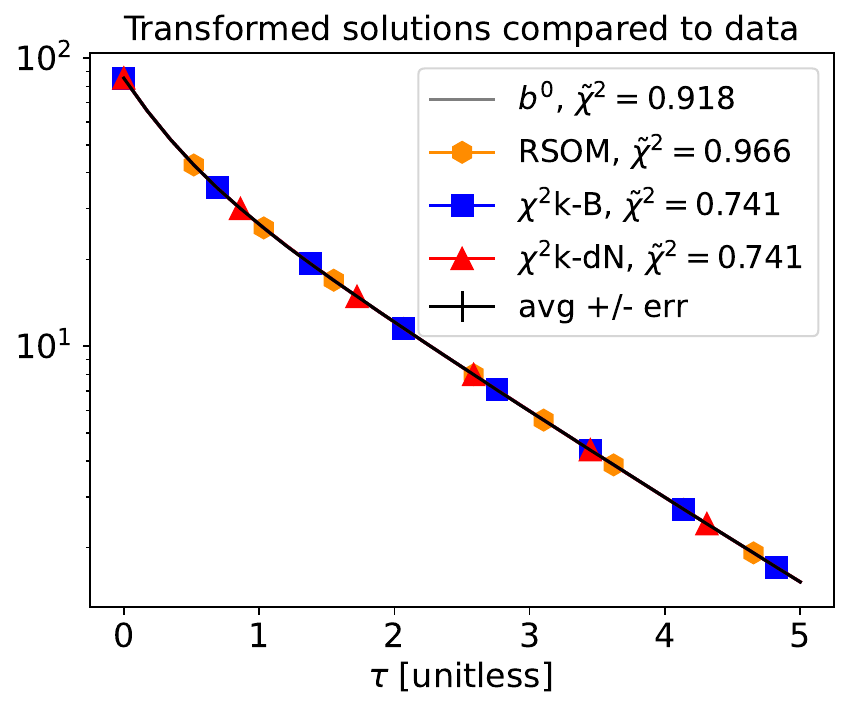}
    \caption{This plot demonstrates the competitive performance of the RSOM in comparison to entropic methods (which incorporate a Bayesian prior) for the double Gaussian test. Left: Solutions in $\omega$-space. Right: Transformed solutions in $\tau$-space with their associated goodness-of-fit to the data (\textit{i.e.}, the reduced chi-square).}
    \label{fig:doubleGaussian}
\end{figure}

Here we consider the double Gaussian test problem, (\textit{i.e.}, problem 1 from Goulko \textit{et al}.~\cite{Goulko_PRB_2017}), which consistently confounds neural network, stochastic, and entropic -based approaches when noise is finite~\cite{wang_PRD_2022, ShaoSandvik_PhysRep_2023, chuna_PRR_2026}. In this example, the RSOM outperforms entropic approaches, but still has room for improvement. This test problem uses the Laplace transformation
\begin{align}
    \mathcal{A}(\tau,\omega) = e^{-\tau \omega} \, ,
\end{align}
formulated on uniform grids $\omega_k \in [4.0/N_\omega, 4.0 ]$ with $N_\omega =150$ and $\tau_i \in [0, 5]$ with $N_\tau = 30$. We create a noiseless signal by transforming the true solution $\xv^0$, a two Gaussian sum,
\begin{align}
    x^0_k = \sum_{i=1}^2 \cfrac{c_i}{\sigma_i} e^{-\cfrac{(\omega_k-z_i)^2}{2\sigma_i^2}} \, ,
\end{align}
via $A_{ik}= \Delta \omega \, T(\tau_i,\omega_k)$. Here the parameters are, $c_1=0.62; \sigma_1=0.12; z_1=0.74$ and $c_2=0.41; \sigma_2=0.064; z_2=2.93$, so that the narrow Gaussian is at large $\omega$. Plots of $\xv^0$ and $\bv^0$ are given in Figure~\ref{fig:doubleGaussian} left and right.
From $b_0$, we generate $N_{\mathrm{s}}=100$ samples by adding Gaussian noise that is scaled by the $\bv^0$ to each component
\begin{align}
    b^s_i \sim \, b^0_i + \mathcal{N}\left(z = 0, \sigma = b^0_i \sigma_0 \right),
\end{align} 
here $\sigma_0 = 0.01$. The mean and variance of these $N_{\mathrm{s}}$ measurements defines the input data $b=\langle A x \rangle$ and the noise covariance $C = \mathrm{diag}(\text{Var}[Ax])/N_{\mathrm{s}}$. We create the regression matrix for the RSOM algorithm using matrix multiplication as $R_{ij} = \sum_k A_{ik} K_{jk}$, where $K_{jk}(\omega) = \exp[-(\omega_k - \mu_j)^2/\sigma^2_j]$. 

Again we compare RSOM to the common entropic least squares method. We find that RSOM greatly outperforms entropic methods, see Figure~\ref{fig:doubleGaussian}. As in the $\rho$-meson problem, it is important that Bryan's approximate optimizer and the dual Newton optimizer yield the same solution. While the RSOM performs well, the fit is not perfect. It may be expected that since the RSOM is fitting two-Gaussians to data created from two-Gaussians that the fit ought to be perfect. However, this is not the case and is readily explained by the particular structure of the Laplace transform, which is often understood as a low-pass band filter in signal  processing~\cite{istratov_RewSciIns_1999}. Essentially, high $\omega$ signals decay rapidly so they do not contribute to the data. In Figure~\ref{fig:doubleGaussian}, we can see the second peak only contributes meaningfully for $\tau < 0.5$, so $>90\%$ of the data informs the first peak. Correspondingly, the RSOM's first Gaussian peak is visually indistinguishable from the true solution.

\subsection{Synthetic data from Laplace transformed skewed Gaussian spectrum}\label{app:skewgaussian}
\begin{figure}
    \centering
    \includegraphics[width=0.5\linewidth]{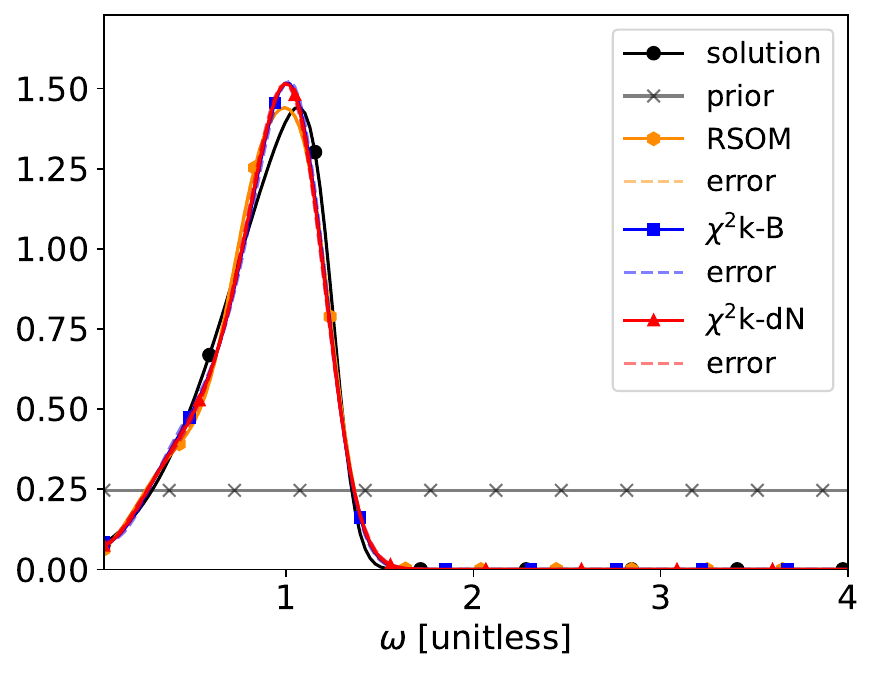}%
    \includegraphics[width=0.5\linewidth]{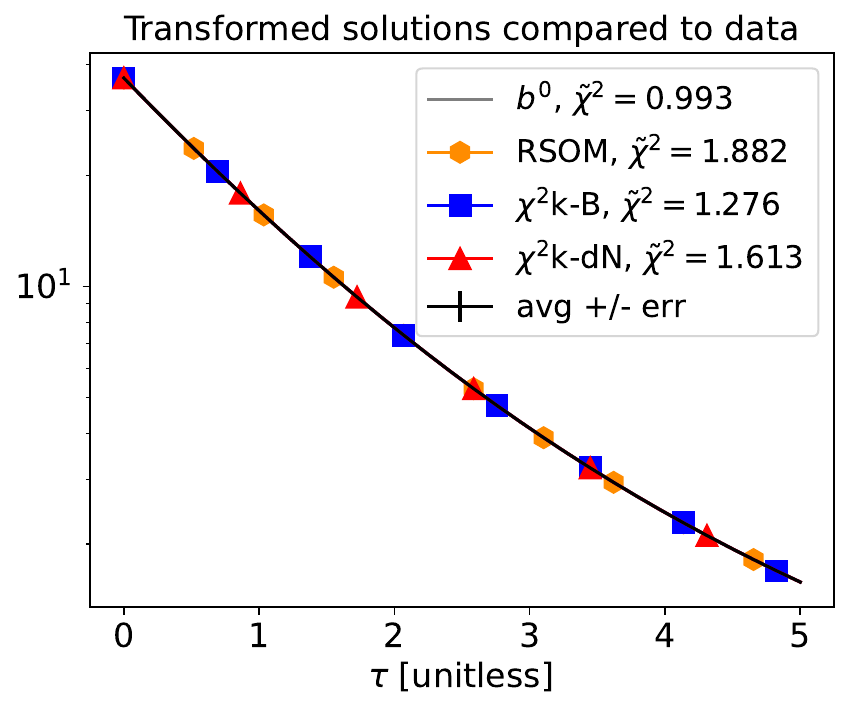}
    \caption{This plot demonstrates RSOM's performance on the AC of the skew Gaussian test is comparable to that of entropic methods which incorporate a Bayesian prior. Left: Solutions in $\omega$-space. Right: Transformed solutions in $\tau$-space with their associated goodness-of-fit to the data (\textit{i.e.}, the reduced chi-square).}
    \label{fig:skewGaussian}
\end{figure}

Skewed functions present a difficulty for the RSOM because Gaussians are symmetric. Here we consider such a problem, generating synthetic data generated from the Laplace transform of a skewed Gaussian spectrum. Skewed Gaussians occur commonly in finite temperate electron systems, which will be the subject of section~\ref{sec:UEG}. In this problem we use the same transformation $\mathcal{A}$, $\tau$-grid and $\omega$-grid as in the double Gaussian problem, but create a noiseless signal by transforming the true solution $\xv^0$, a skewed Gaussian,
\begin{align}
x^0_k &= c_1 \, \frac{2}{\sigma_1}\, \mathrm{PDF(t_k)}\cdot \mathrm{CDF}(t_k) \, ,
\end{align}
which is constructed from a normal probability density function (PDF) multiplied by a cumulative distribution function (CDF) to introduce asymmetry. 
\begin{align}
t &= \frac{\omega - z_1}{\sigma_1},
\\ \mathrm{PDF}(t) &= \frac{1}{\sqrt{2\pi}} \exp\left(-\frac{t^2}{2}\right),
\\ \mathrm{CDF}(t) &= \frac{1}{2}\left[1 + \mathrm{erf}\left(\frac{a_1 \,  t}{\sqrt{2}}\right)\right].
\end{align}
Here the parameters are, normalization $c_1=1.0$, width $\sigma_1=0.5$, center $z_1=1.25$, and skewness $a_1=-5$. The negative skew shifts weight toward higher frequencies. From $b_0$, we generate $N_{\mathrm{s}}=100$ samples by adding Gaussian noise that is scaled by the $\bv^0$ to each component
\begin{align}
    b^s_i \sim \, b^0_i + \mathcal{N}(z = 0, \sigma = \sigma_0),
\end{align} 
here $\sigma_0 = 0.001$. Plots of $\xv^0$ and $\bv^0$ are given in Figure~\ref{fig:skewGaussian}.

We present the solutions obtained on the skewed Gaussian problem for the entropic least squares and for the RSOM in Figure~\ref{fig:skewGaussian}. Comparing the solutions we find that the entropic least squares approach and the RSOM method all perform exceptionally well. This result reinforces that the RSOM is capable of constructing the smooth structures we expect to find in the electron gas. 

For this test problem, the dimensionality scan determines that three Gaussians are needed to sufficiently represent a single skewed Gaussian. Compared to the other examples, the skewed Gaussian has the slowest reduction in the $\chi^2$ with respect to increasing the number of Gaussians. For further discussion of the dimensionality scan process for the skew Gaussian problem, see Figure~\ref{fig:DL_2} located in~\ref{app:implementation}.

\subsection{Synthetic data from Gaussian-smeared sinusoidal spectrum}
\begin{figure}
    \centering
    \includegraphics[width=0.5\linewidth]{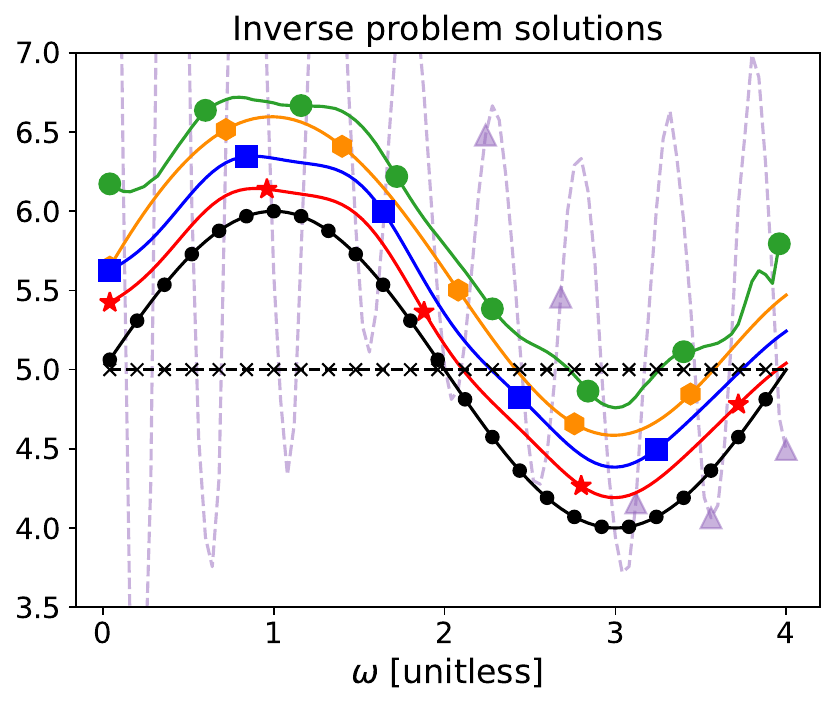}%
    \includegraphics[width=0.5\linewidth]{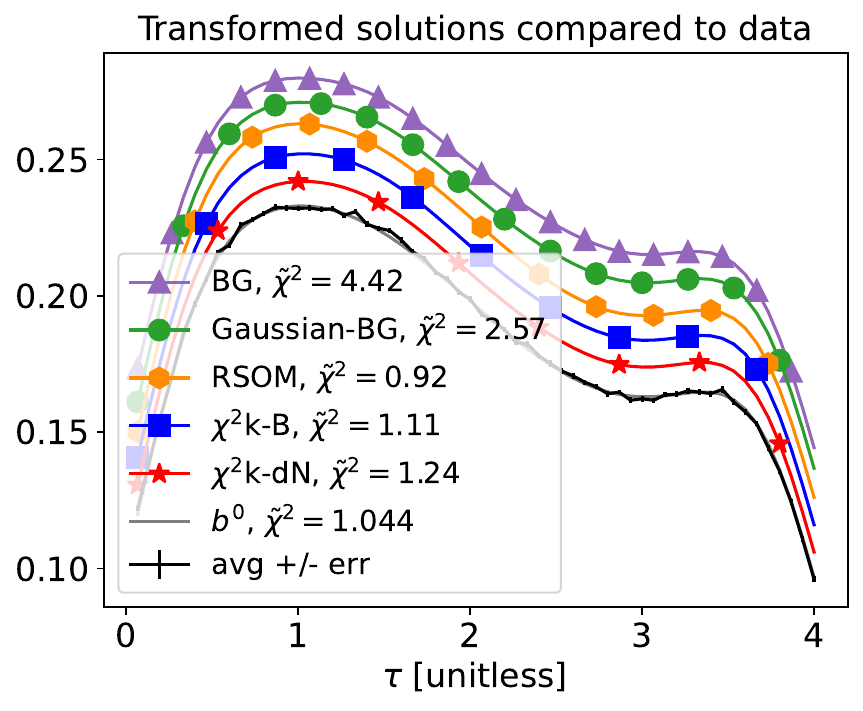}
    \caption{This plot compares the RSOM's results for a Gaussian transformed sinusoidal signal are competitive with other all-purpose inverse methods. Only the entropic methods rely on the Bayesian prior. Left: Solutions in $\omega$-space. We have made the chaotic Backus-Gilbert curve transparent for visual clarity. Right: Transformed solutions in $\tau$-space with their associated goodness-of-fit to the data (\textit{i.e.}, the reduced chi-square). Notice, A small displacement has been added to distinguish between the six curves.}
    \label{fig:GaussianKernel}
\end{figure}

To demonstrate the generality of RSOM, we provide an example of a general ill-conditioned inverse problem. We consider a Gaussian smearing transformation,
\begin{align}
    \mathcal{A}(\tau,\omega)=\exp\left(-\frac{1}{2 \sigma_1}(\tau-\omega)^2\right)\, .
\end{align}
The $\sigma_1$ parameter determines the conditioning of the problem; see that $\sigma_1 \rightarrow 0$ yields the identity matrix and $\sigma_1 \rightarrow \infty$ yields a singular matrix of $1$'s. We use $\sigma_1 = 0.1$ with uniform grids $\omega_k \in [4/N_\omega,4]$ with $N_\omega=100$ and $\tau_i \in [4/N_\tau,4]$ with $N_\tau=60$ to compute matrix $A_{ik}=T(\tau_i,\omega_k)$. We create a noiseless signal by transforming the true solution $\xv^0$, a sinusoidal wave with an offset
\begin{align}
    x^0_k = 5 + \sin\left(\frac{\pi}{2}\omega_k \right) \, ,
\end{align}
resulting in $\bv^0 = A \xv^0$; notice that $\xv^0$ is strictly positive. Essentially, the rectangular matrix $A$ conducts Gaussian smearing with neighboring values of $x$. Plots of $\xv^0$ and $\bv^0$, which are given in Figure~\ref{fig:GaussianKernel} left and right, demonstrate this behavior. From $b_0$, we generate $N_{\mathrm{s}}=100$ samples by adding Gaussian noise to each component, 
\begin{align}
    b^s_i \sim \, b^0_i + \mathcal{N}(\mu = 0, \sigma = \sigma_0) \, ,
\end{align}
where $\sigma_0=10^{-2}$. The mean and error of these $N_{\mathrm{s}}$ measurements defines the input data $b=\langle A x \rangle$ and the covariance matrix $C = \mathrm{diag}(\text{Var}[Ax])/N_{\mathrm{s}}$. We create the regression matrix for the RSOM algorithm using matrix multiplication as $R_{ij} = \sum_k A_{ik} K_{jk}$, where $K_{jk}(\omega) = \exp[-(\omega_k - z_j)^2/\sigma^2_j]$. 

To assess the RSOM, we compare its estimate to four other methods that are commonly used in the literature: (1) the traditional Backus-Gilbert method from Numerical Recipes~\cite{press_NumericalRecipes_2007}, (2) its recent Gaussian variant~\cite{Hansen_PRD-GBG_2019}, which is similar in spirit to Gaussian process regression~\cite{DelDebbio_EuroPhysC_2025}, (3) the entropic regularization approximated via Bryan's modified Levenberg-Marquardt optimizer~\cite{bryan_EuroBiophys_1990, Jarrell_PhysRep_1996, Asakawa_PPNP_2001, rothkopf_Data_2020} (4) as well as the full entropic regularization solved via the dual Newton optimizer~\cite{chuna_JPA_2025}. For the entropic methods, which have a regularization weight, we use to the $\chi^2$-kink procedure~\cite{Kaufmann_CPC-anacont_2023}.

Figure.~\ref{fig:GaussianKernel}, shows the solution for each of these methods, as well as the transformed solution alongside the data. We find that the Backus-Gilbert algorithm performs the worst, while the Gaussian Backus-Gilbert improves significantly over its predecessor. Yet, still both methods under-perform entropic approaches. Typically, entropic methods handle the regularization weight via Bayesian posterior averaging~\cite{gull_MaxEntBayesianMethods_1989, bryan_EuroBiophys_1990}. However, we found that the solutions were under-regularized and switched to the $\chi^2$-kink method~\cite{Kaufmann_CPC-anacont_2023}. In comparison to these entropic results, the RSOM performs similarly, but is slightly smoother. In short, this test demonstrates that the RSOM is competitive with other general inverse methods. 

\section{Application of the RSOM to authentic QMC data from electron gas system}\label{sec:UEG}

In the previous sections we demonstrated the RSOM on synthetic problems where the solution is known. In this final numerical results section, we  investigate authentic ITCFs obtained from QMC simulations of the finite temperature electron gas. No exact solution is known, but this system has been extensively studied by analytic continuation methods~\cite{Filinov_PRB_2023, chuna_JPA_2025, chuna_PRB_2025, chuna_arXiv_2026, dornheim_PRL-DLFC_2018}. The data used in this section is available~\cite{HZDR_datarepository}.

In this system, the analytic continuation is given as
\begin{subequations}\label{eq:periodicLaplace}
\begin{align}
    F(q,\tau) &= \int_{-\infty}^\infty d\omega\ e^{-\tau\omega} S(q,\omega) 
    \\ &=  \int_0^{\infty} d\omega \left( e^{-\tau \omega} + e^{-(\beta - \tau) \omega} \right) S(q,\omega)\ ,
\end{align}
\end{subequations}
where the second equality follows from the detailed balance condition $S(q,-\omega)=S(q,\omega)e^{-\beta\omega}$. Here $S$ is the dynamic structure factor (DSF) and $F$ is the density-density imaginary-time correlation function (ITCF); we rely on the same ITCF data used in~\cite{chuna_PRB_2025}. Only the diagonal of the covariance matrix is used, assuming that the data is independent across $\tau$. The variance is computed using Hatano's error analysis~\cite{hatano1994data} and verified with leave-one-out-binning~\cite{berg_book_2004}. 

The DSF is related to the system's dynamic linear density response function $\chi(q,\omega)$, where $q$ and $\omega$ are the wavenumber and frequency of a weak external harmonic perturbation, via the fluctuation dissipation theorem~\cite{GiulianiVignale_quantumtheory_2008},
\begin{align} \label{eq:FDT}
    S(q,\omega) =  - \frac{1}{ \pi n} \frac{\text{Im} \chi(q,\omega)}{1-e^{- \beta \hbar \omega} } \, ,
\end{align}
where $n$ is the number density. Without loss of generality, the density response is often expressed as~\cite{ichimaru2018plasmavol1, ichimaru1982stronglycoupledplasma, kugler_JStatPhys-LFC_1975} 
\begin{align}\label{eq:susceptibility}
    \chi(q,\omega) = \frac{\chi^0(q,\omega)}{1 - v(q) \left[1 - G(q,\omega)\right] \chi^0(q,\omega)}.
\end{align}
Here $\chi^0(q, \omega)$ describes the density response of the non-interacting electron gas at finite temperatures, $v(q)$ is the Coulomb interaction between two electrons, and $G(q,\omega)$ is the local field correction that alters the interaction to include many-body, \textit{i.e.}, exchange-correlation effects. 

In the weakly-coupled limit the contribution of the many-body term $G$ to the shape of the DSF is small. In this limit, the static approximation $G(q,\omega)=G(q,\omega=0)$~\cite{dornheim_JCP-MLstatic_2019, dornheim_PRB-ESA_2021}, derived from density response theory~\cite{GiulianiVignale_quantumtheory_2008, Dornheim_review}, performs well~\cite{Filinov_PRB_2023, chuna_PRB_2025, dornheim_PRL-DLFC_2018, chuna_JCP_2025}. Investigations have even found that the approximation captures many important physical behaviors outside of the weakly coupled limit. As such, we take the static approximation as the reference solution for an electron gas system with Wigner-seitz radius $r_s=2$ and degeneracy parameter $\Theta = T / E_F =1$. 

\begin{figure}
    \centering
    \includegraphics[width=0.75\linewidth]{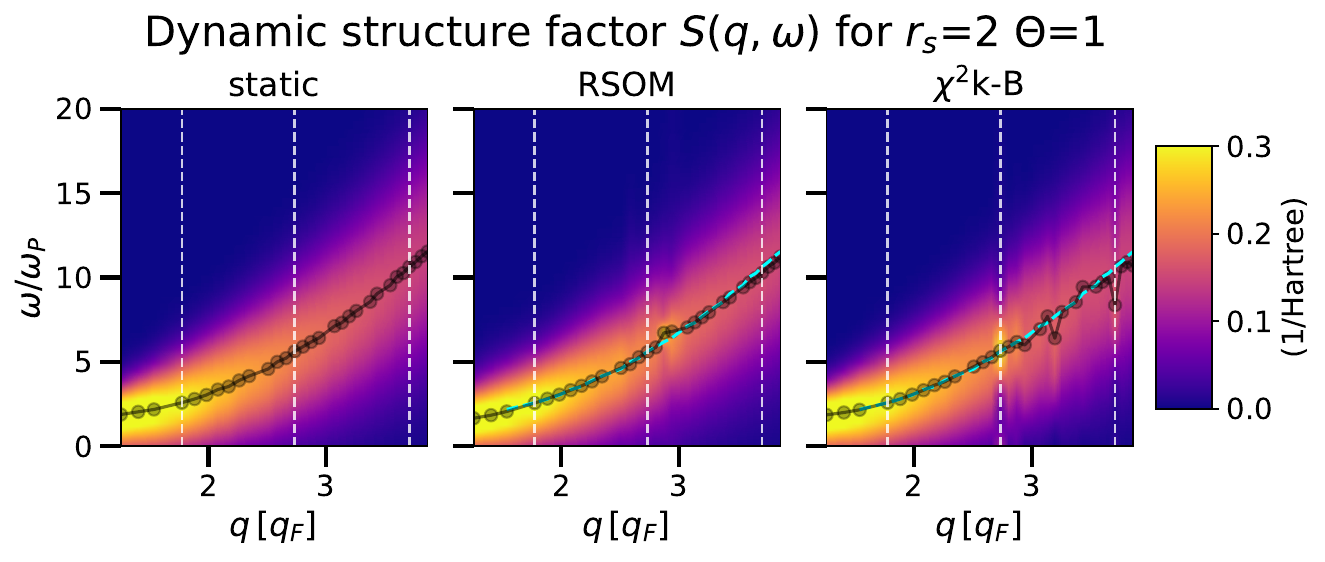}
    \includegraphics[width=\linewidth]{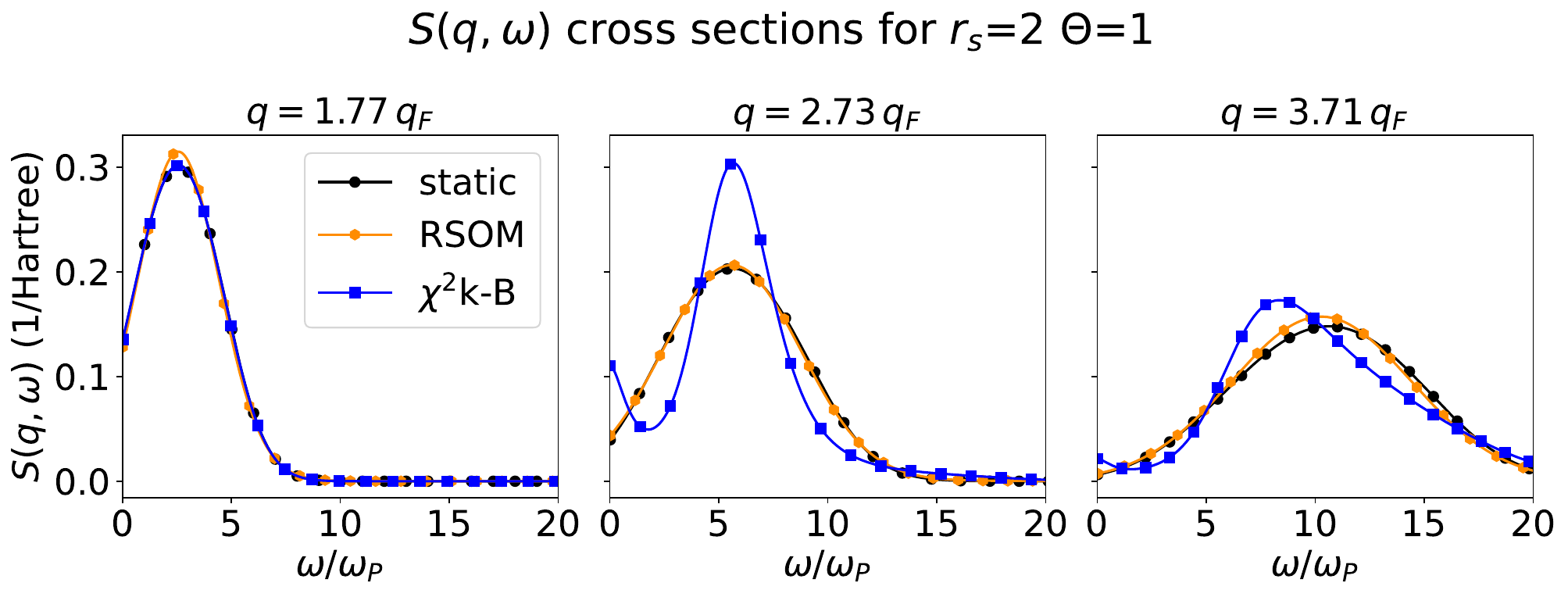}
    \caption{Top: Heatmaps presenting the analytic continuation results for the dynamic structure factor of the finite temperature electron gas for $N=34$ particles at Wigner-seitz radius $r_s=2$ and degeneracy parameter $\thetav = T / E_F =1$; frequencies are given in units of the plasma frequency $\omega_\textnormal{p}=\sqrt{3/r_s^3}$ and wavenumbers in units of the Fermi wavenumber $q_\textnormal{F}=(9\pi/4)^{1/3}/r_s$. Bottom: Cross sections of the heatmap taken where the white lines indicate. The static approximation is very reliable in this parameter region, often understood as the true answer. The entropic $\chi^2$k-B has been given the static approximation as its prior.}
    \label{fig:UEGDSF}
\end{figure}

In Figure~\ref{fig:UEGDSF}, we compare the static approximation (\textit{i.e.}, the expected solution), the RSOM estimate, and the entropic least squares estimate. The RSOM produces more stable results than the entropic least squares at large wavenumbers $q$. We highlight the dispersion relation, as assessed by the location of the DSF's maximum value and indicated by black dots on the heatmaps, which is more stable at large $q$ with respect to the entropic least squares. This stability is an important achievement because the entropic least squares has the static approximation as its prior, while RSOM does not rely on the static approximation. The entropic least squares' instability at large $q$, even with an exceptionally good prior, is a known problem~\cite{chuna_JCP_2025} and follows from the rapidly decaying ITCF data, which has few appreciable data points that are not statistically zero. 

Additionally, in Figure~\ref{fig:UEGDSF}, we take cross sections from the heatmap above and highlight the worst entropic results. These plots demonstrate that Bryan's entropic least squares optimizer can overfit the data, even though the method tends to be overly smooth~\cite{chuna_JPA_2025, rothkopf_Data_2020} and the regularization parameter was selected via the $\chi^2$kink algorithm (which was originally formulated to prevent over fitting!). For comparison, the entropic least squares estimates typically have $\tilde{\chi}^2 \approx 10^{-6}$, while the RSOM estimates have $\tilde{\chi} \approx 10^{-1}$. As a result of the goodness-of-fit values, which are all less than 1, all methods are expected to satisfy sum rules; this is verified in appendix~\ref{app:sumrules}.

In conclusion, this authentic problem serves as a reassuring application of the RSOM because the data are authentic, the expected result can be estimated through a forward problem, RSOM consistently selects solutions in-line with that expectation, and the estimates satisfy known sum rules.

\section{Quantifying the coefficients' sensitivity to perturbations}\label{sec:conditioning}

Above we demonstrated that the results of the RSOM algorithm are competitive with other methods. While such comparisons are valuable for understanding a method's systematic behavior, they are not a quantitative assessment of the overall solution's sensitivity to perturbations. We address a piece of this problem, by investigating how perturbations in the coefficients' $\delta \cv$ depend on perturbations in the matrix $\delta R$ (for a fixed $N_c$) and the data $\delta \bv$; these quantities are related by the conditioning of the problem $\kappa(R)$. In subsection~\ref{sec:conditioning_def}, we define the conditioning and discuss its limitations, \textit{i.e.}, it neglects the uncertainty in $N_c$ or the meta-choice to use Gaussians. Then in subsection~\ref{sec:conditioning_estimates}, we provide estimates of the conditioning, and finally in subsection~\ref{sec:sensitivity_estimates}, we state the error bound that relates $\delta \cv$ to $\delta R$ and $\delta \bv$ via the condition number, for a fixed $N_c$ Gaussians.

\subsection{Defining conditioning}\label{sec:conditioning_def}
The \textit{continuous} AC problem is well-established as an exponentially ill-posed inverse problem and recent works in both physics and numerical analysis have characterized the singular value structure of AC transformation kernels~\cite{shi_CPC_2023, ohzeki_arXiv_2026, misawa_arXiv_2026}. Related to the posedness, but different, is the conditioning of the discrete problem~\cite{Hofmann_arXiv_2026}, which arises after an $\omega$-grid is selected. Here the dictionary $K(N_c,\thetav)$ plays the role of the $\omega$-grid, formulating the discrete optimization problem \eqref{eq:reformulatedACproblem_discrete} from the continuous AC problem \eqref{eq:ACproblem_continuous}, so the conditioning analysis is carried out on $R$.

The conditioning of the discrete problem associated with either \eqref{eq:ACproblem_discrete} or \eqref{eq:reformulatedACproblem_discrete} is typically quantified via the ratio of the largest and smallest condition number of the discrete matrix to be inverted. For example, in \eqref{eq:ACproblem_discrete} these singular vectors come from matrix $A$ and in \eqref{eq:reformulatedACproblem_discrete} these singular values come from the matrix $R$. This definition of condition number as the ratio of singular values aligns with the definition broadly used in numerical analysis, which is the ratio of the smoothness and strong convexity constants for the discretized loss, often denoted $S_L$ and $\mu_L$. These constants characterize the Hessian of the loss function. For quadratic convex optimization problems (\textit{e.g.} least squares inversion of \eqref{eq:reformulatedACproblem_discrete}) the Hessian is $R^T R$, the smoothness constant is $\sigma_{\max}^2$, and the strong convexity constant is $\sigma_{\min}^2$. As such, the conditioning analysis in the AC literature is the square-root of what is broadly used.

Our choice to investigate the conditioning of $R$ aligns with other AC practitioners who also investigate the conditioning after the discrete problem has been formulated. For example, sparsity promoting methods~\cite{Otsuki_PRE_2017, Otsuki_JPSJ_2020, Motoyama_2022_PRB, ohzeki_arXiv_2026, misawa_arXiv_2026} formulate a new inversion matrix $S$ using intermediate representation (IR) basis vectors and conduct extensive conditioning analysis on the singular values of $S$, and not the matrix $A$ obtained from the typical uniform discretization. As another example, Robles \textit{et al}.~\cite{robles_CPC_2025} formulate a new inversion matrix $R$ using Gaussian basis vectors and also conduct investigations of the singular values structure of $R$. 

We indicate that typical conditioning analyses will not account for the full uncertainty in $K(N_c, \thetav)$, which plays the role of the $\omega$-grid in other formulations. At present we only incorporate the uncertainty in the $R$ for a given $N_c$ Gaussians. This is because discovering the best dictionary $K(N_c, \thetav)$ is a non-convex problem and the notion of conditioning is not well-defined. Essentially, conditioning estimates are for a single unique Hessian, but the RSOM has $N_c = 1, ..., \infty$ different convex quadratic problems and each leads to a different Hessian. Beyond the issue of many minima, there is also the question of which functions make the best dictionary. To assess the full difficulty of selecting $K(N_c, \thetav)$, dictionary learning frameworks are needed, but such analyses have yet to be introduced or discussed in the AC literature.



\subsection{Estimating conditioning of $R$} \label{sec:conditioning_estimates}

For the RSOM, the condition number of the discovered discrete inverse problem is given by the regression matrix $\kappa(R=AK)$. Since stochastic sampling creates a variety of $R$ matrices, as discussed in~\ref{app:implementation}, we collect and average the $\kappa(R)$ associated with all accepted solutions; the averages are presented in Table~\ref{tab:condition-number}. For comparison we have also collected the numerical estimate of the conditioning for the traditional matrix $A$, note that this conditioning estimate is limited by numerical precision, the actual conditioning number is likely much larger~\cite{Hofmann_arXiv_2026}.
\begin{table}[]
    \centering
    \caption{A tabulation of the condition numbers for the synthetic problems studied in this work. The first column indicates the problem name, the second column indicates the conditioning of the traditional formulation $\kappa(A)$, and the remaining columns indicate the conditioning of the discovered discrete inverse problem (\textit{i.e.}, $\kappa(R = A \, K(N_c,\thetav))$) averaged across accepted parameters $\thetav$ at each dimension $N_c$; estimate error is indicated by $()$. The $N_c$ selected by the dimensionality scan algorithm is indicated by a bold value. The selection process used to determine the appropriate number of kernels is discussed in~\ref{app:implementation} and visualized for each problem in Figure~\ref{fig:DL_1}.}
    \begin{tabular}{c | c | c c c c}
        problem & $\kappa(A)$ & $N_c=1$ & $N_c=2$ & $N_c=3$ & $N_c=4$ \\ \hline
        $\rho$-meson & $1.8 \times 10^{19}$ & $1(0)$ & $\mathbf{102(1)}$ & $213(1)$ & $6500(83)$  \\
        2 Gaussians &  $2.0 \times 10^{17}$ & $1(0)$ & $\mathbf{3(0)}$ & $200(\infty)$ & $2700(\infty)$  \\
        skew Gaussian & $2.0 \times 10^{17}$ & $1(0)$ & $22(1)$ & $\mathbf{68(0)}$ & $3600(\infty)$  \\
        sin wave & $1.0 \times 10^{17}$ & $1(0)$ & $\mathbf{3(0)}$ & $16(2)$ & $52(21)$ \\
    \end{tabular}
    \label{tab:condition-number}
\end{table}

We see from Table~\ref{tab:condition-number}, that for each step in the dictionary learning algorithm's dimensional scan from $N_c=1,...,4$ the conditioning becomes much worse, but in practice only a few Gaussians are needed to represent the data. We find that the conditioning of the discovered discrete inverse problem is negligibly small compared to the original problem, (\textit{i.e.}, $\kappa(R) \ll \kappa(A)$), as can be seen from Table~\ref{tab:condition-number}. In practice we find that after the critical dimension selected by the dimensionality scan, additional Gaussians can strongly overlap with each other and this leads to singular matrices. We remove these matrices to compute the average, but present an infinite standard error to indicate this.

\subsection{Quantifying the coefficients' sensitivity to perturbations}\label{sec:sensitivity_estimates}

Here we consider the error in solving for $\cv$ within \eqref{eq:RSOMopt}, but not the full error. An upper bound on the relative error in RSOM's coefficients $\Vert \delta c \Vert / \Vert c \Vert$ is given from linear perturbations of the solution with respect to the data $b$ and the matrix $R$ to be inverted. For condition number $\kappa(R)$, the relative error in the data $\| \delta F \| / \| F \|$, and the relative error in the matrix $\| \delta R \| / \| R \|$ the bound is given~\cite[Theorem 7.2]{Higham_textbook_2002}
\begin{align}\label{eq:OLS_MSE}
    \cfrac{ \Vert \delta c \Vert}{\Vert c \Vert} 
    & \leq \cfrac{\kappa(R)}{1 - \kappa(R)  \cfrac{\Vert \delta R \Vert}{\Vert R \Vert} } \left( \cfrac{\Vert \delta b \Vert}{\Vert b \Vert} + \cfrac{\Vert \delta R \Vert}{\Vert R \Vert} \right) \, . 
\end{align}
The correction factor in the denominator arises from the cross dependencies between uncertainty in the matrix and the solution, \textit{i.e.}, the term $\delta R \, \delta c$ is preserved in the perturbation analysis. The denominator must be positive, which we find to be the case in our examples, but this fails for large $N_c$ where both the condition number and the relative uncertainty increase. This formula does not account for uncertainty in the choice of Gaussian or the number of Gaussians on $\| \delta c \| / \| c \|$.

The quantities in \eqref{eq:OLS_MSE} can be computed from either the data or the stochastic samples as: 
\begin{align}
    \| b \| &= \| \textrm{Avg} \left( b \right) \|_2 \, ,  &&  \| \delta b \| = \left\| \sqrt{ \frac{\textrm{Var} (b)}{ N_\textrm{data}}} \right\|_2
\\    \| R \| &= \| \textrm{Avg}_\textrm{sto} \left( R \right) \|_2 \, ,  && \| \delta R \| =\left\| \sqrt{ \frac{\textrm{Var}_\textrm{sto} ( R ) }{ N_\textrm{sto}}} \right\|_2 \, . 
\end{align}
Here $\textrm{sto}$ indicates that the average and variance are taken with respect to RSOM's stochastic samples.
We use $N_{sto}$ to account for the fact that stochastic sampling does not generate independent samples, this is the number of collected samples normalized by the integrated autocorrelation time (IACT) $\tau_\mathrm{IACT}$.
In short, $\| \delta b \|$ is the 2-norm of the standard error in the data vector, while $\| \delta R \|$ is the 2-norm (largest singular value) of the standard error in the $R$ matrix. Empirically, when flipping the order of avg and 2-norm the result varies less than $< 5\%$. 

As a heuristic, we propose that $\Vert \delta c \Vert / \Vert c \Vert < 1\%$ indicates the coefficients are stable, while $\Vert \delta c \Vert / \Vert c \Vert < 10\%$ indicates a moderate sensitivity. Lastly, if $ \kappa(R) \times  \Vert \delta b \Vert / \Vert b \Vert \approx 1$, then there is possibly $100\%$ relative error, so the method should be considered to have failed. In Table~\ref{tab:problem-statistics} column 5, we see that the estimated upper bound is small for the skew gaussian and sin-wave synthetic problems, but is large for the double Gaussian problem. This coincidentally matches an intuitive understanding of when the RSOM is performing well or performing poorly.
\begin{table}[]
    \centering
    \renewcommand{\arraystretch}{1.25}
    \setlength{\tabcolsep}{12pt}
    \caption{For the synthetic problems studied in this work, we tabulate the condition numbers for the typical formulation $\kappa(A)$ and the RSOM's discovered discrete inverse problem $\kappa(R)$. This comparison establishes the vastly improved conditioning of the discovered discrete AC problem. We also present the relative norm of the error $\Vert \delta b \Vert / \Vert b \Vert$, and compute the upper bound on the relative error in the solution coefficients $\Vert \delta c \Vert / \Vert c \Vert$ using \eqref{eq:OLS_MSE}.}
    \begin{tabular}{c | c | c | c | c | c }
        problem & $\kappa(R)$ & $ \nicefrac{\Vert \delta \bv \Vert}{\Vert \bv \Vert}$ & $ \nicefrac{\Vert \delta R \Vert}{\Vert R \Vert}$ & $ \leq  \nicefrac{\Vert \delta \cv \Vert}{\Vert \cv \Vert}$ \\ \hline
        $\rho$-meson &   $102$ & $4.7 \times 10^{-4}$ & $1.6 \times 10^{-4}$& $\le 13 \%$ \\
        2 Gaussians & $3$ & $1 \times 10^{-3}$ & $5.3 \times 10^{-2}$ & $\le 20\%$ \\
        skew Gaussian & $68$ & $7 \times 10^{-6}$ & $1\times 10^{-4}$ & $\le 1\%$ \\
        sin wave & $3$ & $5.3 \times 10^{-3}$ & $2\times 10^{-4}$ & $\le 2 \%$\\
    \end{tabular}
    \label{tab:problem-statistics}
\end{table}

Note that the RSOM selects a single Gaussian for the authentic QMC data, so the conditioning is guaranteed $\kappa(R)=1$ and the sensitivity estimate is low because the data has a small relative error $< 10^{-3}$ and the uncertainty in the Gaussian kernel is also small.

\section{Summary and conclusions} \label{sec:conclusions}

In summary, we have formulated, implemented, and demonstrated the regularized stochastic optimization method (RSOM) for analytic continuation. For the problems considered here, the RSOM discovers a sparse Gaussian dictionary that formulates a low-dimensional discrete AC problem. This optimization problem has finite uncertainty and achieves a reduced chi-square $\tilde{\chi}^2 < 2$. The dictionary learning formulation presented in \eqref{eq:DLformulation} recovers the conventional AC problem when the sparse Gaussians are replaced with many Dirac deltas on a fixed dense $\omega$-grid. As such, the RSOM is formally leveraging the dictionary that is present in every discretized AC problem. 

From a practical perspective, the RSOM is suited for multi-peak problems, as is demonstrated by its competitive performance on the double Gaussian test problem, which has broadly confounded AC methods. Furthermore, the RSOM performs well on smooth structures. This is demonstrated by the $\rho$-meson and skewed Gaussian test problems. Furthermore, the RSOM is a general inverse problem solver and can be applied to any AC problem. This is verified by its application to the Gaussian-smeared sinusoidal. Finally we note that, in its application to authentic data, the RSOM remains stable in the large $q$ limit of the finite temperature electron gas, thus it mitigates a known challenge. 

\textit{We draw two major conclusions from this work:}

Firstly, though the origin of the ill-posedness in the continuous AC problem is known~\cite{misawa_arXiv_2026}, the ill-conditioning associated with typical uniform grid discretization can be mitigated by generalizing the $\omega$-grid to a kernel set (\textit{i.e.}, dictionary) and then learning a sparse dictionary from the data. This parsimonious regularization exchanges the ill-conditioned and under-constrained optimization problem for the NP-hard dictionary learning problem and for the cases considered here can lead to finite uncertainty in the dictionary parameters and coefficients.

Secondly, the hyper-parameter engineering commonly done in both the stochastic and regularized AC communities can be understood as kernel-set engineering (\textit{e.g.}, $\omega$ grid engineering~\cite{bergeron_PRE_2016, chuna_arXiv_2026, beach_arXiv_2004, ghanem_PhDthesis_2017, ShaoSandvik_PhysRep_2023}). Therefore, we suggest that, in the spirit of sparsity, the coarsest $\omega$-grid that can represent the data well should be used. In practice, if the grid is not tuned to the data, then it ought to be selected based on a Bayesian prior. A few interesting studies already examine the impact of the grid~\cite{Ghanem_PRB_2020, richardson_arXiv_2025, ohzeki_arXiv_2026} and an overview of sampling theorems related to the Laplace transform are given by Istratov and Vyvenko~\cite{istratov_RewSciIns_1999}. 

Future work might improve upon the implementation of \eqref{eq:RSOMopt}, which parses the AC problem into a brute-force dimension scan and two nested low-dimensional optimizations. There is no reason this is the best implementation of the method. Other non-smooth optimization algorithms could be used to determine parameters $\thetav$~\cite{clason_arXiv_2026} and other dictionary learning algorithms are presented in the compressed sensing literature~\cite{davenport_compressedsensing_2012, foucart_compressedsensing_2013} as well as emerging ``unrolled'' dictionary learning \cite{Monga_IEEE_2021, Mathelin_ACME_2018, tolooshams_arXiv_2021, malezieux_arXiv_2021, malezieux_thesis_2023}. 

Future work might also consider that a major limitation of the RSOM is that although Gaussians are universal function approximators, are local in $\omega$-space, and have well-defined moments, they are a structural prior and introduce approximation bias. This is particularly important for certain functional forms that require many Gaussians to represent. Since we have seen that the condition number grows with $N_c$, increasing $N_c$ will eventually lead to an ill-conditioned inverse problem. As such, practitioners must inform their dictionary choice with prior knowledge. The approximation theory literature provides alternatives to the Gaussian dictionary~\cite{Trefethen_SIAM_2019, Weeks_JACM_1966} as does the analytic continuation literature, which has recently begun exploring the intermediate representation (IR) basis~\cite{ohzeki_arXiv_2026, misawa_arXiv_2026, Motoyama_2022_PRB}. We emphasize an older study by Davies and Martin~\cite{Davies_JCompPhys_1979}, that compared many different dictionaries of a fixed size (\textit{e.g.} Laguerre and Legendre polynomials). While they hand-tuned dictionaries without dictionary learning algorithms, the expansive list of dictionaries provides inspiration.

Lastly, future work ought to investigate, compare, and improve upon the coefficient error bound proposed in section~\ref{sec:conditioning}. For example, investigate the robustness to data using resampling procedures, assess the impact of the dimensionality scan using non-convex optimization tools, compare across dictionary functions using tools such as the common AIC (Akaike Information Criterion) or BIC (Bayesian Information Criterion), or less common spark number or coherence~\cite[chapter 1]{foucart_compressedsensing_2013}; many options exist.




\section*{Acknowledgments}

TC dedicates his efforts in this work to his high school physics teacher Jeff McManus, who is retiring this year after 33 years of contributions to his students, workplace, and community. Thank you, Mr. McManus, for being an excellent educator and more.

TC acknowledges Nicholas J. E. Richardson and Michael P. Friedlander for insightful discussion on optimization algorithm convergence as well as adaptive $\omega$-grid algorithms. ChatGPT was used to check for grammar and clarity issues and Claude suggested the use of a Langevin proposal as a way to improve sampling efficiency in the Metropolis-Hasting algorithm.

This work has received funding from the European Union's Just Transition Fund (JTF) within the project \emph{R\"ontgenlaser-Optimierung der Laserfusion} (ROLF), contract number 5086999001, co-financed by the Saxon state government out of the State budget approved by the Saxon State Parliament. This work has received funding from the European Research Council (ERC) under the European Union’s Horizon 2022 research and innovation programme (Grant agreement No. 101076233, "PREXTREME"). Views and opinions expressed are, however, those of the authors only and do not necessarily reflect those of the European Union or the European Research Council Executive Agency. Neither the European Union nor the granting authority can be held responsible for them. TD gratefully acknowledges funding from the Deutsche Forschungsgemeinschaft (DFG) via project DO 2670/1-1. This work
was supported by the Center for Advanced Systems Understanding (CASUS), which is financed by Germany’s
Federal Ministry of Research, Technology and Space
(BMFTR) and by the Saxon State government out of the
State budget approved by the Saxon State Parliament.

Computations were performed on a Bull Cluster at the Center for Information Services and High-Performance Computing (ZIH) at Technische Universit\"at Dresden and at the Norddeutscher Verbund f\"ur Hoch- und H\"ochstleistungsrechnen (HLRN) under grant mvp00024.





\bibliography{bibliography}

\appendix

\section{Implementation}\label{app:implementation}
In this appendix we describe the implementation of \eqref{eq:RSOMopt}. For convenience, the equation is reproduced the equation here, 
\begin{align}
    S(\omega) = \underset{N_c}{\text{scan}} \left[ \min_{\thetav} \left( \min_{\cv \ge 0} \chi^2[K(N_c,\thetav), \, \cv] \right) \right] \, .
\end{align}
As noted this problem could be solved by many different algorithms, but here the choice is made to combine a dimensionality scan, and gradient-informed Metropolis-Hastings (MH) sampling for kernel parameter $\thetav$ optimization, and scipy's~\cite{SciPy2020NMeth} standard non-negative least squares (NNLS) optimization for the kernel coefficient $\cv$ estimation. Our code is freely available online~\cite{github_MEMcode}.

\subsection{Dimensionality Scan}
In our implementation, the dimensionality scan is the top-level routine. It collects solutions for various $N_c$ and then makes an assessment based on the $\chi^2$ values associated to each $N_c$. This brute-force search is the simplest of dictionary learning methods, yet has been successfully used in a few physics applications~\cite{chuna_phdthesis_2024, ellison_aps-dpp_2018, svensson_arXiv_2026}.

In the dictionary learning literature, it has been observed that as the dimension $N_c$ of the dictionary increases there is typically some critical dimension, where the $\chi^2$ drops dramatically. This criteria is often used to select the dimension $N_c$. However, in this work, we must also be sure that the fit has a sufficiently high goodness-of-fit. So in the event that the chi square has a large drop, but its value is not less than 2, we continue increasing the dimension until $\chi^2 < 2$, \textit{i.e.}, the data are well represented.

\begin{figure}
    \centering
    \includegraphics[width=0.49\linewidth]{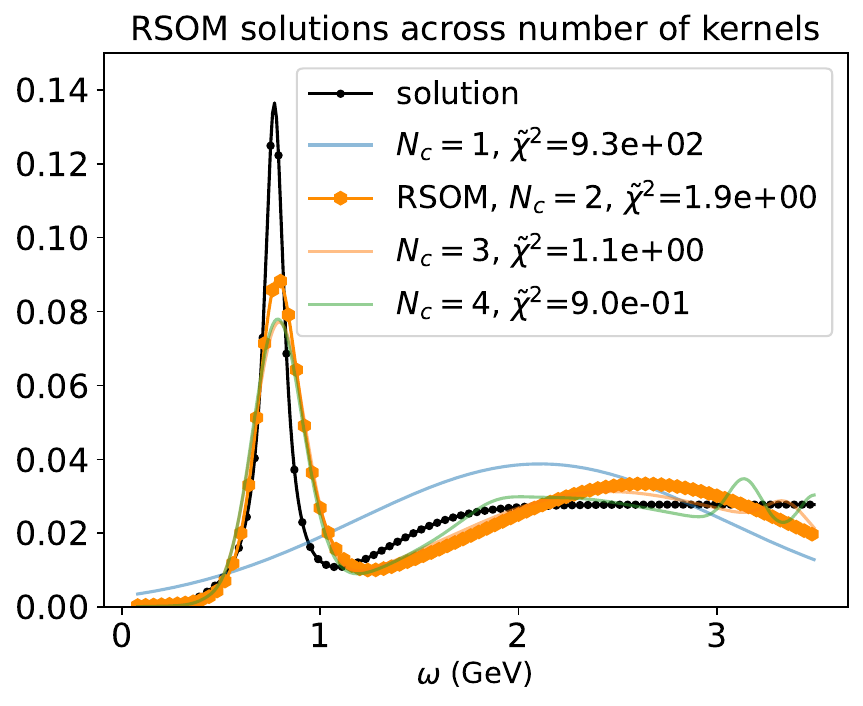}%
    \includegraphics[width=0.51\linewidth]{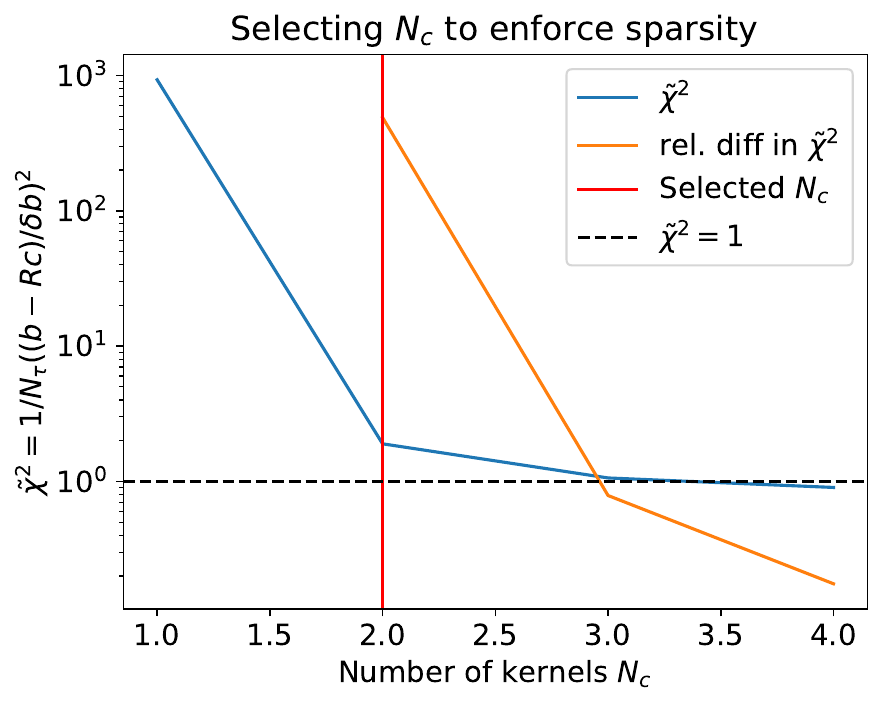}
    \includegraphics[width=0.47\linewidth]{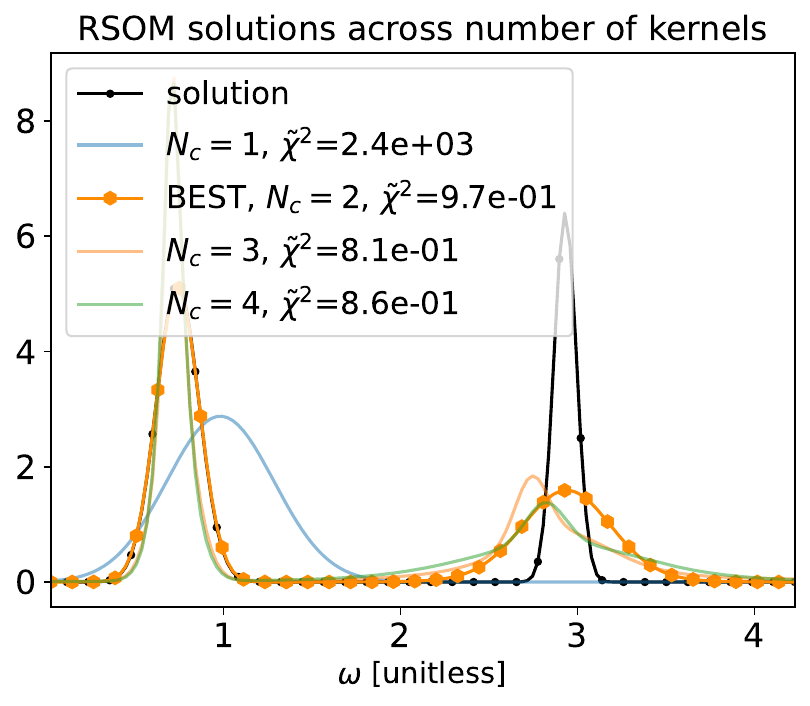}%
    \includegraphics[width=0.53\linewidth]{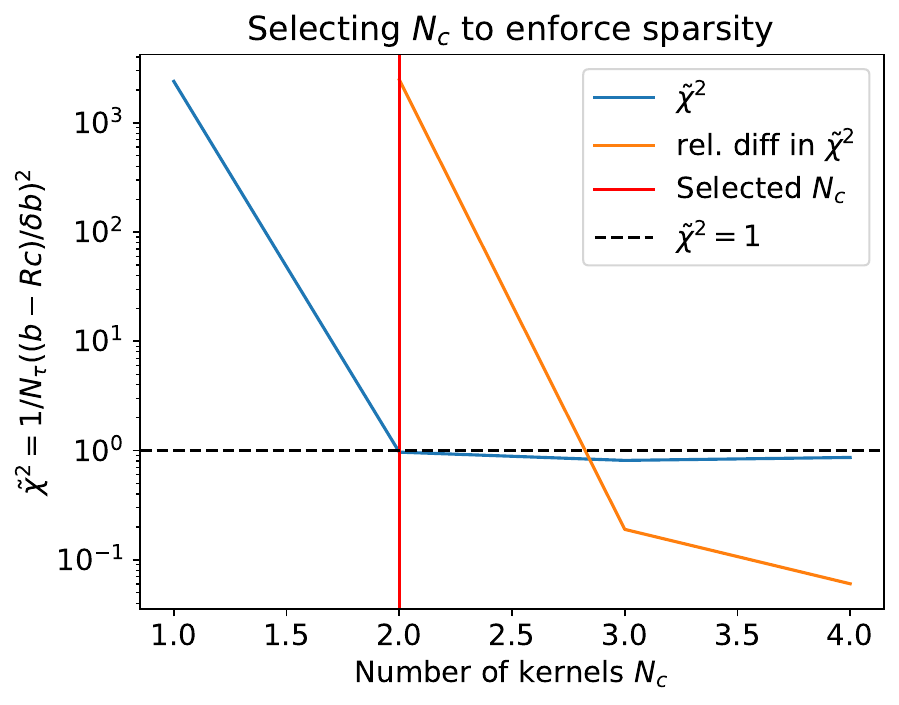}
    \caption{This plot visualizes how the RSOM's dimensional scan algorithm, (\textit{i.e.}, selecting the number of Gaussians needed to represent the solution), as it is applied to the $\rho$-meson test problem (top row) and the double Gaussian test problem (bottom row). The left column presents the average solution obtained for each step in the dimensionality scan, so $N_c$ indicates the number of Gaussians in the dictionary. The right column presents the reduced chi-square $\tilde{\chi}^2 = \chi^2/N_\tau$ as well as the relative difference between steps in the dimensionality scan. The red line indicates the number of Gaussians selected by the dictionary learning algorithm.}
    \label{fig:DL_1}
\end{figure}

\begin{figure}
    \centering
    \includegraphics[width=0.48\linewidth]{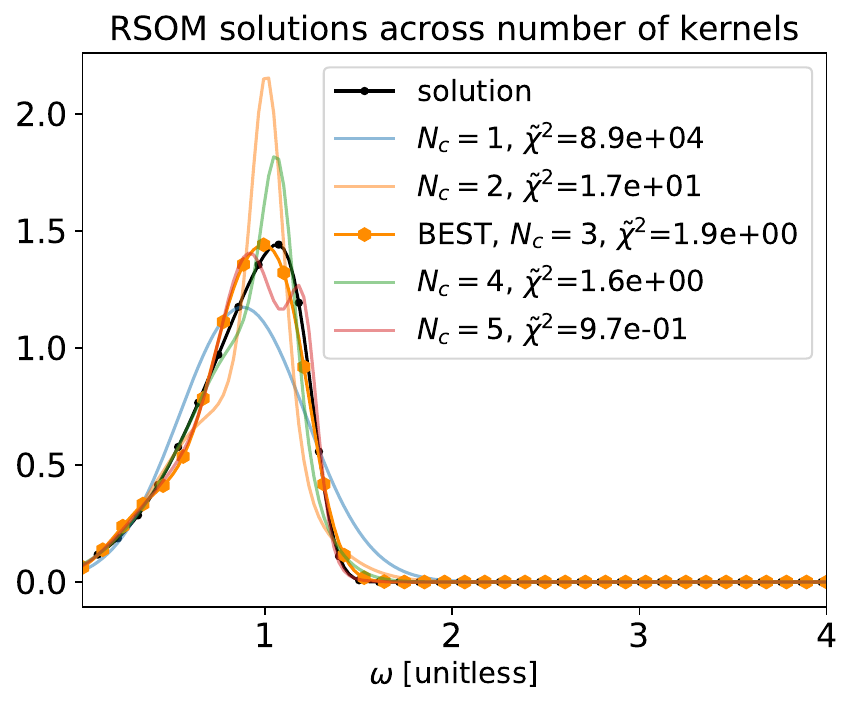}%
    \includegraphics[width=0.52\linewidth]{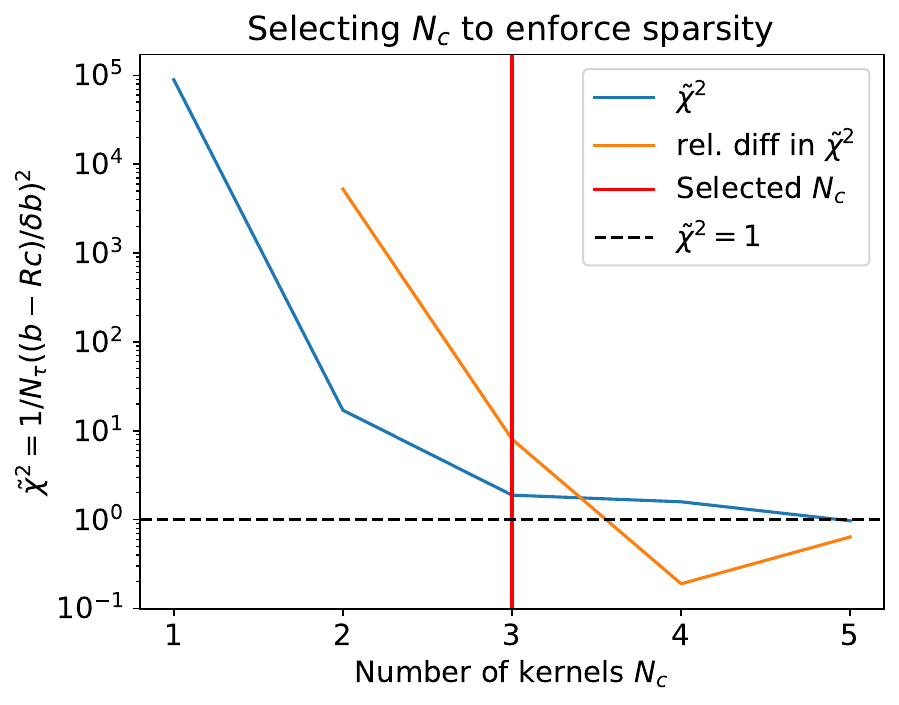}
    \includegraphics[width=0.48\linewidth]{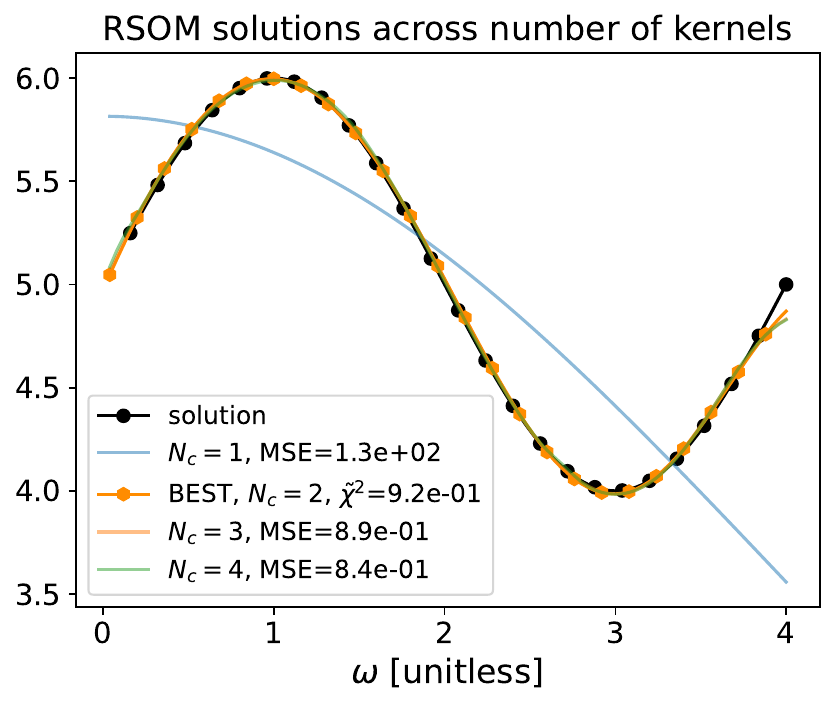}%
    \includegraphics[width=0.52\linewidth]{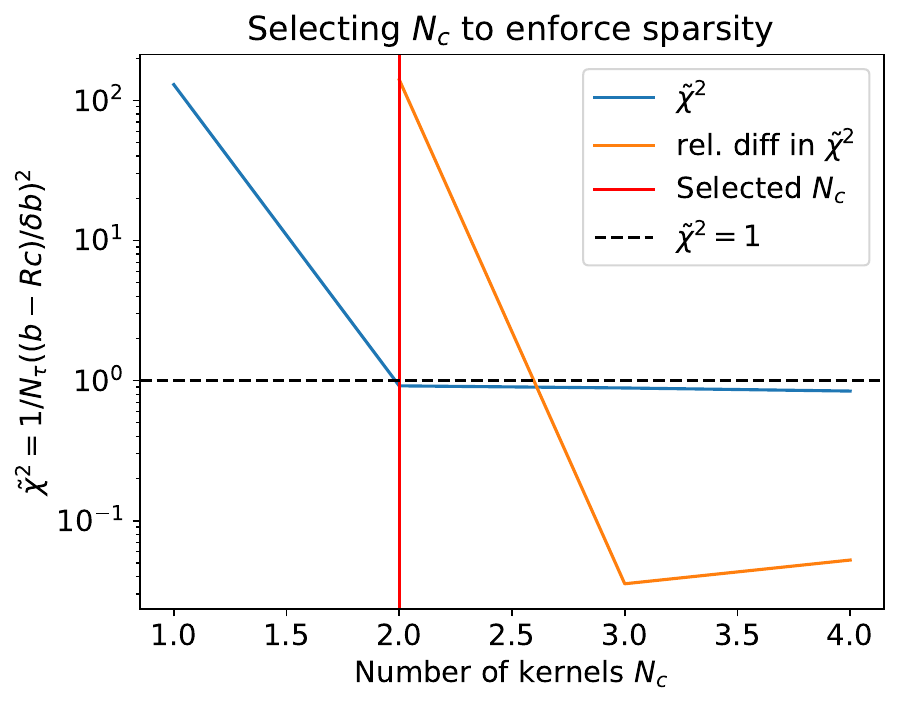}
    \caption{This plot visualizes how the RSOM's dimensional scan algorithm, (\textit{i.e.}, selecting the number of Gaussians needed to represent the solution), as it is applied to the Laplace transformed skewed Gaussian (top row), and the Gaussian-smeared sinusoidal (bottom row). The left column presents the average solution obtained for each step in the dimensionality scan, so $N_c$ indicates the number of Gaussians in the dictionary. The right column presents the reduced chi-square $\tilde{\chi}^2 = \chi^2/N_\tau$ as well as the relative difference between steps in the dimensionality scan. The red line indicates the number of Gaussians selected by the dictionary learning algorithm.}
    \label{fig:DL_2}
\end{figure}

We provide pseudocode of our dimensionality scan in algorithm~\ref{alg:DL} and visualize the selection procedure in Figure~\ref{fig:DL_1} and Figure~\ref{fig:DL_2} for the synthetic problems from section~\ref{sec:results}. In general, we find that $N_c < 5$ works, and thus these problems need only a few Gaussians.

\begin{algorithm}
\caption{Dimensionality scan pseudocode}
\label{alg:DL}
\begin{algorithmic}[1]
\Require
\Statex
$\begin{aligned}
&A_0, \bv_0, C, \omegav, \tauv \\
&N_{\min}, N_{\max}
\end{aligned}$

\State Compute Cholesky whitening of transformation matrix $A_0$ and data $b_0$ using data covariance matrix $C$.

\Statex
\For{$N_c = N_{\min}$ to $N_{\max}$}
    \State Sample solution space and compute the avg. and var. of the solution.
    \begin{align}
        (x_{\text{RSOM}}, \sigma^2_{\text{RSOM}}) \gets \texttt{stoch\_opt}(A,\bv,\omegav,\tauv,N_c)
    \end{align}

    \State Compute reduced chi-square of average solution:
    \begin{align}
    \chi^2 = \frac{\|A\xv_{\text{RSOM}} - \bv\|_2^2}{N_\tau}
    \end{align}

    \State Store average solution and its variance.
\EndFor

\State Select optimal kernel count from maximum:
\begin{itemize}
    \item first $N$ where $\chi^2 < 2$
    \item $N$ with largest relative drop in $\chi^2$
\end{itemize}

\State Return selected solution and uncertainty.

\end{algorithmic}
\end{algorithm}

In Figure~\ref{fig:DL_1}, we consider the $\rho$-meson and the double Gaussian problems. Interestingly, Asakawa \textit{et al}. claimed in their seminal work~\cite{Asakawa_PPNP_2001} that a $\rho$-meson spectral function is similar to a narrow and a broad Gaussian combination, an intuition that we confirm. For the double Gaussian problem, the RSOM unsurprisingly identifies that two Gaussians are present. 

Moving to Figure~\ref{fig:DL_2}, we present the sinusoidal wave and the skewed Gaussian. We find the full period of the sine wave is represented very well with only two Gaussians. Essentially the two Gaussians straddle the trough. The skewed Gaussian requires the largest number of Gaussians to be represented. This is unsurprising since a Gaussian is symmetric and many Gaussians of diminishing magnitude are needed to represent a long tail. Notice that the $\tilde{\chi}^2$ for the skew Gaussian test problem has the most gradual reduction in $\chi^2$ value with respect to $N_c$, \textit{i.e.}, the other three test problems have a very abrupt drop at $N_c=2$.  

As shown, the dimensionality scan serves as the parent function in the code, looping over the dimension $N_c$ to gather solutions, and then assessing them relative to each other. The function calls a subfunction stoch\_opt to gather those solutions. In the next section we will discuss this subfunction in detail.

\subsection{Minimization over dictionary parameters}
We now focus on the inner minimizations over Gaussian parameters $\thetav$ and coefficients $\cv$, presented in \eqref{eq:RSOMopt}. These optimizations could be handled simultaneously. However, the optimization over $\thetav$ is non-linear, while the optimization over $\cv$ is linear. Further, the conditioning degrades as the number of Gaussians increases. As such, we use a Metropolis-Hastings (MH) sampling routine on $\thetav$ to mitigate this degradation and a linear routine on $\cv$ to reduce the cost. Since the minimization over $\cv$ is handled by the scipy~\cite{SciPy2020NMeth} package, this section primarily focuses on the minimization over $\thetav$.

\begin{figure}[h]
    \centering
    \includegraphics[width=0.75\linewidth]{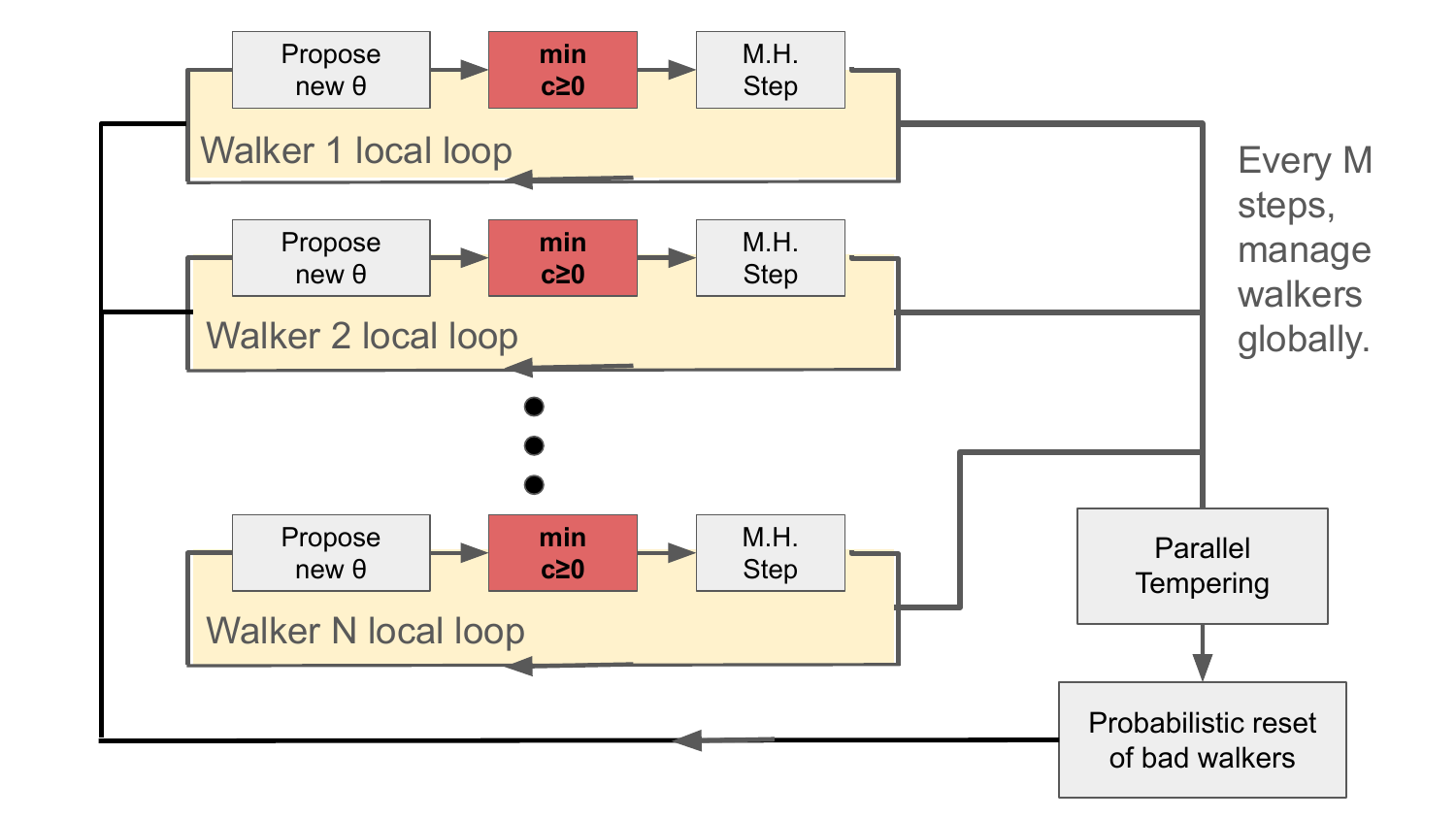}
    \caption{A schematic of the RSOM stochastic optimization of the Gaussian parameters $\thetav$. Each walker executes the following loop. First, it proposes a new set of parameters $\thetav$. Second, it solves a non-negative least squares optimization to find the best fit of $\cv$ to the data ($\cv \geq 0$). Third, it assesses whether the update improved the fit of the data according to the Metropolis-Hastings (M.H.) criteria. After each walker conducts local updates, the algorithm manages the walkers globally using parallel tempering and genetic resets to improve sampling efficiency.}
    \label{fig:RSOMschematic}
\end{figure}

The RSOM MH sampling routine uses many walkers to explore the space of Gaussian parameters. Each walker generates new proposals for $\thetav$ and accepts/rejects based on the $\chi^2$ achieved by the inner $\min_{\cv \ge 0} \chi^2$, so every proposal requires a fast non-negative least squares (NNLS) algorithm to solve for $\cv$. Each walker has a different MH temperature, so they accept and reject with different probabilities. The algorithm leverages this population via global updates. First, parallel tempering assigns high-performing walkers colder MH temperatures, so they are less likely to move away from their local minimum. Second, genetic resets re-initialize low-performing walkers. This global management, while breaking ergodicity, substantially improves sampling efficiency. A visual schematic of this routine is provided in Figure~\ref{fig:RSOMschematic}.

In practice, the MH sampling algorithm collects solutions after a burn-in period. After a sufficiently large set of solutions has been collected, the best solution and its uncertainty are obtained by computing the average and variance of this set. The associated pseudocode is given in Algorithm~\ref{alg:stoch_opt}. 

\begin{algorithm}[h!]
\caption{Stochastic optimization pseudocode}
\label{alg:stoch_opt}
\begin{algorithmic}[1]

\Require
\Statex
$\begin{aligned}
&\mathrm{Whitened \,\, system} \, (A,\bv), \omegav,\tauv, N_c
\end{aligned}$

\Statex

\State Initialize parallel tempering temperatures.
\State Initialize $N_\beta$ Walkers each with $N_c$ Gaussians.
\begin{align}
&z_i = \mathcal{U}[\, \omega_{\min} , \, 0.95 \omega_{\max} \,] \\
&\sigma_i = \mathcal{U}[\, 2 \Delta \omega, \, 0.8 \, (\omega_{\max} - \omega_{\min} ) \, ] \\
&\thetav = \left(\zv , \sigmav \right) 
\end{align}





\For{$i = 1$ to $N_{\mathrm{PT}} + N_{\mathrm{burn\_in}}$}

    \For{each walker}

        \For{$m=1$ to $N_{\mathrm{MH}}$}

            \State Langevin propose $N_c$ new Gaussians:
            \begin{align}
                \thetav' \gets \texttt{langevin\_proposal} (\thetav)
            \end{align}

            \State Construct associated matrices:
            \begin{align}
                &K \gets \texttt{kernel\_matrix}(\omega,\tau, \thetav')
                \\
                &R = A\,K
            \end{align}

            \State Solve nonnegative least squares:
            \begin{align}
                \cv = \arg\min_{\cv \ge 0} \|R\,\cv-\bv\|_2^2
            \end{align}
            
            \State Compute change in chi-square.
            \begin{align}
                \Delta \chi^2 = \chi^2(\thetav')-\chi^2(\thetav)
            \end{align}

            \State Metropolis-Hastings acceptance:
            \If{$u < \exp(-\beta \, \Delta \chi^2)$ }
                \State Accept proposal.
                \If{$i > N_{\mathrm{burn\_in}}$ and $\chi'^2 < 2 \chi^2_{\min}$ }
                    \State Store $\xv = K \, \cv$
                \EndIf
            \EndIf

        \EndFor
    \EndFor

    \State Perform parallel tempering swaps between neighboring walkers.
    \State Randomly respawn poorly performing walkers.

\EndFor

\State Compute integrated autocorrelation time $\tau_\mathrm{int}$ for $N_{\mathrm{eff}}=  N_\mathrm{samples} / \tau_\mathrm{int}$.

\State Return mean and variance of collection solutions.
\end{algorithmic}
\end{algorithm}

A peculiarity of the RSOM's MH implementation is that we exploit the fact that there are relatively few parameters $\thetav$ to be optimized. Thus, instead of a uniformly random proposal, the MH algorithm uses a gradient-informed proposal, termed a Langevin proposal. Our numerical implementation is given in Algorithm~\ref{alg:langevin}, where $\epsilon$ is an \textit{ad-hoc} parameter to adjust the noise.
\begin{algorithm}\label{alg:langevin}
\caption{langevin proposal pseudocode}
\begin{algorithmic}[1]
\Require Current state $\thetav$

\State Compute $\chi^2$ gradient, $\gv = \nabla_{\thetav} \chi^2$
\State Compute $\chi^2$ Hessian, $H = \nabla_{\thetav} \nabla_{\thetav'} \chi^2$
\State Compute Cholesky factor of the Hessian $L$

\State Compute adaptive step scaling .
\State Compute drift direction:
\[
H^{-1} \gv = L^{-T}L^{-1} \gv
\]

\State Compute stochastic noise:
\[
\eta \sim \mathcal{N}(0,H^{-1})
\]

\State Form Langevin step:
\[
\Delta \thetav = - H^{-1} \gv + \sqrt{\frac{2\epsilon}{\beta}} \etav
\]

\State Return proposal:
\[
\thetav' = \thetav + \Delta\thetav
\]
\end{algorithmic}
\end{algorithm}

In summary, the MH algorithm uses walkers to solve many NNLS optimizations, searching for a low-dimensional space that represents the data using only a small number of kernels. Here, the MH update ensures the walkers improve their kernel set's ability to represent the data, while hedging against the impending ill-posedness that emerges as more Gaussians are included to improve the data representation, \textit{i.e.}, when $N_c$ increases. 

\section{Sum rule verification}\label{app:sumrules}
In this appendix, as an independent assessment of the method, we verify that the RSOM estimates satisfy known sum rules. This indicates that the DSFs describe certain physics. Here, we investigate the inverse sum rule, static structure factor (zeroth sum rule), and the f-sum rule, which are given as:
\begin{align}
    \langle \omega^{-1} \rangle &= - \frac{\chi(q,0)}{2n} = \frac{1}{2} \int_0^\beta d\tau \, F(q,\tau) \label{eq:inverse}\,,
    \\ \langle \omega^{0} \rangle &= S(q) = F(q,\tau=0) \label{eq:Sq}\,,
    \\ \langle \omega \rangle &= E_F \left(\frac{q}{k_F}\right)^2 = - \left. \frac{\partial}{\partial \tau} F(q,\tau) \right\rvert_{\tau=0} \label{eq:f-sum}\, , 
\end{align}
with $\langle \omega^{k} \rangle  \equiv \int d \omega \, \omega^k S(q,\omega)$. The first equality in \eqref{eq:inverse} tells us that the static density response function is sensitive to when the DSF's peak is located at small $\omega$ and is presented in, for example, \cite{Vitali_PRB_2010, dornheim2024QuantumDelocalization}. Thus, systems with a roton feature are expected to have larger density response in the intermediate wavenumbers $q$ where the DSF peak is red-shifted. The second equality, presented \textit{e.g.}, in~\cite{dornheim2024QuantumDelocalization}, tells us that if the integral of the ITCF is small then there is little density response. Equation \eqref{eq:Sq} defines the familiar static structure factor $S(q)$, which gives the system's average spatial correlations~\cite{hansen2013theory3rdEd}. Finally, Equation \eqref{eq:f-sum} defines the f-sum rule, which is often associated with the continuity equation~\cite{boon_molecularhydro_1991, kremp2005quantum}. 

\begin{figure}
    \centering
    \includegraphics[width=0.35\linewidth]{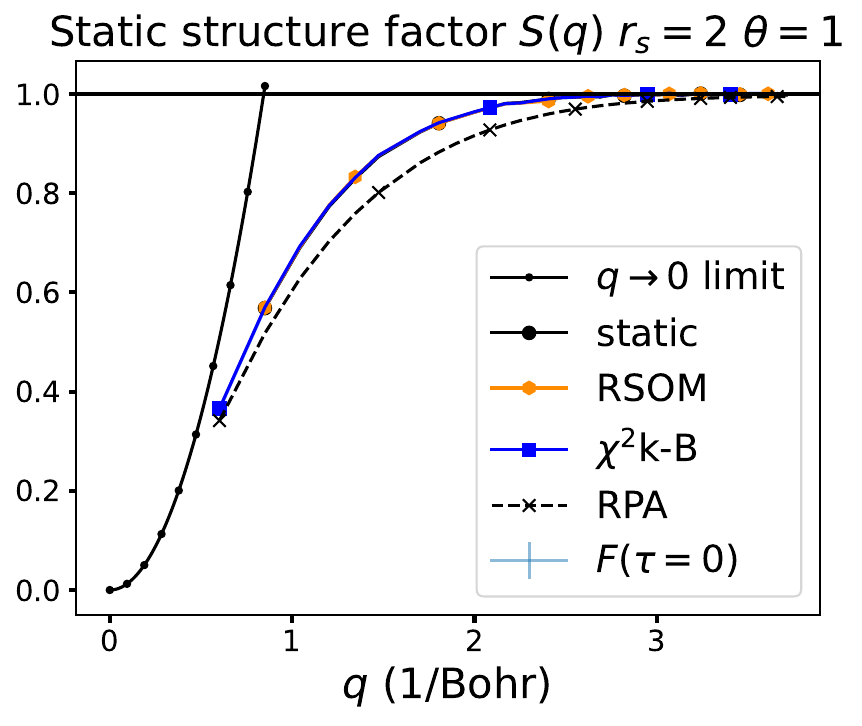}%
    \includegraphics[width=0.30\linewidth]{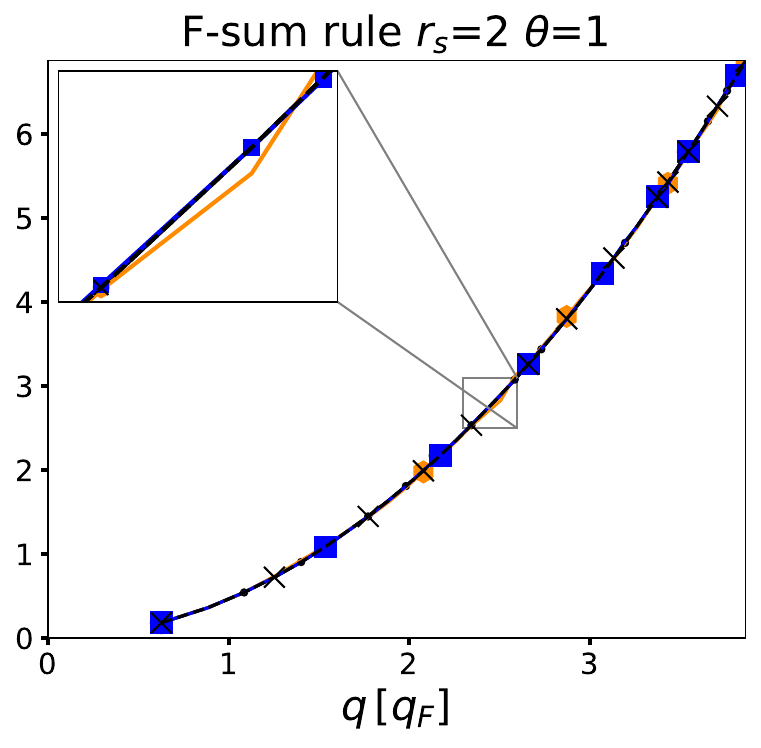}%
    \includegraphics[width=0.35\linewidth]{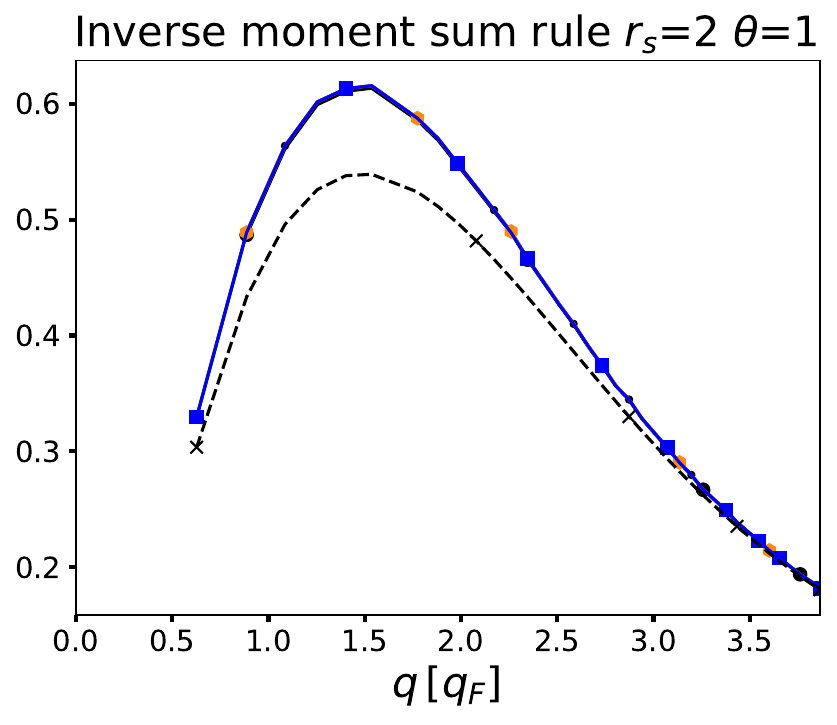}%
    \caption{We plot the static structure factor \eqref{eq:Sq} (left), the f-sum rule \eqref{eq:f-sum} (middle), inverse sum rule \eqref{eq:inverse} (right). The black circles indicate the moments of the static approximation across wavenumber $q$, while the darkorange hexagons and blue squares indicate the typical formulation solved with the RSOM and entropic least squares (with the static as a prior and the $\chi^2$ kink algorithm~\cite{Kaufmann_CPC-anacont_2023} to set the regularization weight). The insets use the same color code and demonstrate a small violation in f-sum satisfaction.}
    \label{fig:UEGDSF_sumrules}
\end{figure}

In Figure~\ref{fig:UEGDSF_sumrules}, we see that the RSOM finds agreement with known sum rules. This agreement is to be expected because sum rules are expressed in terms of derivatives and integrals of the ITCF data. Thus matching the data well, as measured by the reduced chi-square $\tilde{\chi}^2$, is sufficient to satisfy sum rules~\cite{chuna_PRB_2025}. We see in Figure~\ref{fig:UEGDSF_sumrules} that the static approximation also matches all the sum rules.

\end{document}